\documentclass[preprint2]{aastex}

\slugcomment{to appear in ApJ}

\shorttitle{Nucleosynthesis in merged WD binaries}
\shortauthors{Garc\'{\i}a-Hern\'andez et al.}

\begin{document}

\title{CNO abundances of HdC and RCB stars: a view of the
nucleosynthesis in a white dwarf merger}

\author{D. A. Garc\'\i a-Hern\'andez\altaffilmark{1}, K. H. Hinkle\altaffilmark{2}, David. L. Lambert\altaffilmark{3}, K. Eriksson\altaffilmark{4}}
\altaffiltext{1}{Instituto de Astrof\'{\i}sica de Canarias, C/ Via L\'actea
s/n, 38200 La Laguna, Spain; agarcia@iac.es}
\altaffiltext{2}{National Optical Astronomy Observatory (NOAO), Tucson, AZ85726,
USA; hinkle@noao.edu}
\altaffiltext{3}{W. J. McDonald Observatory. The University of Texas at
Austin. 1 University Station, C1400. Austin, TX 78712$-$0259, USA; dll@astro.as.utexas.edu}
\altaffiltext{4}{Department of Physics and Astronomy, Uppsala University,
Box 515, 75120 Uppsala, Sweden; Kjell.Eriksson@astro.uu.se}

\begin{abstract}
We present high-resolution (R$\sim$50,000) observations of  near-IR transitions
of CO and CN of the five known hydrogen-deficient carbon (HdC) stars and four R
Coronae Borealis (RCB) stars. We perform an  abundance analysis of these stars
by using spectrum synthesis and state-of-the-art  MARCS model atmospheres for
cool hydrogen-deficient stars. Our analysis confirms reports by Clayton and
colleagues that those  HdC stars exhibiting CO lines in their spectrum and the
cool RCB star S Aps are strongly enriched in $^{18}$O (with $^{16}$O/$^{18}$O
ratios ranging from 0.3 to 16). Nitrogen and carbon are in the form of $^{14}$N
and $^{12}$C, respectively. Elemental abundances for CNO are obtained from
C\,{\sc i}, C$_2$, CN, and CO lines. Difficulties in deriving the carbon
abundance are discussed. Abundances of Na from Na\,{\sc i} lines and   S from
S\,{\sc i} lines are obtained. Elemental and isotopic CNO abundances suggest
that HdC and RCB stars may be related objects and that they probably formed from
a merger of a He white dwarf with a C-O white dwarf.
\end{abstract}


\section{Introduction}

The category of hydrogen-deficient luminous stars includes, in order of
increasing effective temperature, the hydrogen-deficient carbon (HdC) stars, the
R Coronae Borealis (RCB) stars, and the extreme helium (EHe) stars. A general
supposition is that this sequence represents an evolutionary one. Studies of the
chemical compositions of stars along the sequence provide one tool for
establishing the validity of the claim for an evolutionary sequence and for
testing proposed mechanisms for placing  stars on the sequence.  This study of
high-resolution infrared spectra of  selected HdC and RCB stars  was prompted by
the remarkable discovery from medium-resolution infrared spectra that the oxygen
in HdCs was primarily the isotope $^{18}$O and not the usual dominant isotope
$^{16}$O (Clayton et al. 2005, 2007).

The origins of the HdC,  RCB and EHe stars  have remained a puzzle for decades.
Two different scenarios have survived theoretical and observational scrutiny. In
one, the H-deficient supergiant is formed as a consequence of the merger of a He
white dwarf (WD) with a carbon-oxygen WD (Webbink 1984; Iben \& Tutukov 1984;
Saio \& Jeffery 2002). This is generally referred to as the double-degenerate
(DD) scenario. In the other, these H-deficient stars result from a final,
post-AGB helium shell flash in the central star of a planetary nebula (PN)
(Iben, Tutukov \& Yungelson 1996). This final flash may transform the PN central
star into a cool H-deficient supergiant; the so-called ``born-again'' scenario,
as discussed by, for example, Herwig (2001) and Bl\"ocker (2001). In a version
of this scenario, the remaining H-rich envelope is ingested by the He-shell and
the ensuing nucleosynthesis includes large-scale conversion of H to He. This is
refered to as the final flash (FF) scenario.

Elemental and isotopic abundances for C, N, and O are a powerful tool for
discriminating between HdC and RCB stars formed by the DD and FF scenarios.
Recent studies of the elemental abundances  favor the idea that most HdC and RCB
stars have formed via the DD scenario (e.g., Saio \& Jeffery 2002;  Rao 2005;
Pandey et al. 2006; Clayton et al. 2006). These studies predicted
compositions for the DD scenario based on the assumption that the DD scenario
did not involve nucleosynthesis during or following accretion of the He white
dwarf by the C-O white dwarf. Clayton et al. (2007) suggested with exploratory
calculations that the accretion is very rapid and induces nucleosynthesis that
converts $^{14}$N by $\alpha$-capture to $^{18}$O and results in a high
abundance of $^{18}$O relative to $^{16}$O. In contrast, conditions in the FF
scenario are unlikely to provide for abundant amounts of  $^{18}$O. Guerrero,
Garc\'\i a-Berro \& Isern (2004) report smoothed particle hydrodynamical
simulations of merging white dwarfs that predict nucleosynthesis to occur in
mergers that might result in a H-deficient single star. Saio \& Jeffery (2002)
modeled the merger with a slow accretion rate (10$^{-5}$ $M_\odot$ yr$^{-1}$
versus the 150 $M_\odot$ yr$^{-1}$ invoked by Clayton et al. [2007]) but
followed the structure of the merger product during and following accretion as
the star evolved  from the blue to the red and back to the blue at approximately
constant (high) luminosity. Structural changes caused by He-burning and, if the
white dwarfs were assumed to have a H-shell, H-burning were followed as were the
dredge-ups to the surface. Obviously, nucleosynthesis occurs in a dramatic
fashion when the accreting white dwarf exceeds the Chandrasekhar limit.

In this paper, we present and analyse high-resolution spectra in selected
intervals in the K band showing CO, CN, and other lines for a  selected sample
of HdC and RCB stars. Our  goal is to do an abundance analysis and, in
particular,  to extract the isotopic ratios $^{12}$C/$^{13}$C,
$^{16}$O/$^{17}$O/$^{18}$O and $^{14}$N/$^{15}$N.  Section 2 describes the
near-IR observations.   The  abundance analysis is presented in Section 3. The
results of this  analysis are discussed in the framework of the DD scenario for
the formation of HdC and RCB stars in Section 4 and concluding remarks are
offered in Section 5.

\section{Observations and overview of the spectra}

Our sample is composed of the five known HdC stars  plus four RCB stars.
High-resolution infrared spectroscopic observations were carried out on June 18
2006, February 1 2007, and February 7 2007  using the PHOENIX spectrograph at
Gemini South (Hinkle et al. 2003). The spectra were obtained with the 0.34
arcsec (R$=50,000$) slit at 5 grating tilts in the K-band region each providing
a bandpass of about 0.01 $\mu$m (19.5 cm$^{-1}$) and centered at 2.251, 2.332,
2.343, 2.354 and 2.366 $\mu$m; the latter four tilts provide essentially
complete coverage from 2.327--2.371 $\mu$m. Table 1 lists the spectral regions
covered by our PHOENIX observations for each star in our sample. The observed
spectra were reduced to intensity as a function of wavelength by using standard
tasks in IRAF\footnote{The Image Reduction and Analysis Facility software
package (IRAF) is distributed by the National Optical Astronomy Observatories,
which is operated by the Association of Universities for Research in Astronomy,
Inc., under cooperative agreement with the National Science Foundation.} and the
telluric features were removed with the help of a spectrum of a hot star
observed the same night. The S/N of the observed spectra is always $>$100 at the
continuum. Wavelengths are given for standard air.

We have constructed the spectral energy distributions (SEDs) of the stars in our
sample by using BVRIJHKL photometry available in the literature and the IRAS
flux density at 12 $\mu$m, in order to check if our sample stars display a
significant infrared excess at 2.3 $\mu$m. In the K band, the flux from HdC
stars is from the photosphere; an infrared excess from circumstellar dust does
not contaminate the K band spectra. The photospheric spectrum is also provided
for the RCB star S Aps with an effective temperature $T_{\rm eff} = 5400$ K  and
observed near maximum light. The RCB UW Cen was observed near maximum light but
its flux in the K band is predominantly from circumstellar dust so that the
photospheric spectrum is greatly obscured or diluted. Even were the star's
photosphere observed free of obscuration, the K band spectra would be expected
to be free of molecular lines for this star with $T_{\rm eff} = 7400$ K.
According to magnitude estimates assembled by the American Association of
Variable Star Observers (AAVSO)\footnote{see http://www.aavso.org/}, V854 Cen
was observed during a  decline of several magnitudes in brightness, suggesting
that the circumstellar emission from hot dust probably dominates the observed
spectrum. Even at maximum light, V854 Cen has an infrared excess contributing at
2.3 $\mu$m. As in the case of UW Cen, absent  an infrared excess, the K band
spectrum would not be expected to show molecular lines because $T_{\rm eff} =
6750$ K (Asplund et al. 1998).  Lack of CO and CN lines in the spectrum of
several stars is most probably due to their effective temperatures. Tenenbaum et
al. (2005) report an absence of CO lines from other RCBs with temperatures
$T_{\rm eff} \geq 6500$ K. With $T_{\rm eff} = 7250$K (Asplund et al. 2000), the
lack of molecular lines for Y Mus is not surprising.

Before describing the synthesis of the spectra, an overview of the spectral
regions is provided here beginning  at the shortest wavelengths observed.
Figures 1 and 2 show the wavelength interval 2.246--2.256 $\mu$m. The CN
molecule through its Red System is a contributor across all observed regions but
lines of this molecule are especially well seen in this region because it is
free of contributions from the first-overtone CO bands. The most prominent
$^{12}$C$^{14}$N lines in the synthetic spectra computed for HD 137613 are
marked by the broken red lines. The CN lines are prominent in the spectra of HD
137613, HD 175893, HD 182040, and S Aps but also present in HD 148839 and HD
173409 (Figure 3). A few lines of the C$_2$ Phillips bands 0-2 and 1-3 cross
this interval. Among these 0-2 Q34 at 2.2492 $\mu$m, 1-3 Q8 at 2.2497 $\mu$m and
Q10 at 2.2528 $\mu$m appear either unblended or sufficiently so to be useful in
spectrum syntheses of S Aps, HD 137613, HD 175893 and HD 182040. (0-2 R22 at
2.2488 $\mu$m may also be useful.) The C$_2$ lines are below the detection limit
for HD 148839, Y Mus, UW Cen and V854 Cen, while HD 173409 was not observed
around 2.251 $\mu$m. (In the available longer wavelength regions, C$_2$ Phillips
lines are irretrievably lost among blends with CN or CO lines.) The 2.251 $\mu$m
region because of its lower line density is also one in which atomic lines are
readily seen. These include a Si\,{\sc i} line at 2.2538 $\mu$m, several S\,{\sc
i} lines at 2.250$-$2.256 $\mu$m, and a Fe\,{\sc i} line at 2.2473 $\mu$m.
Another Fe\,{\sc i} line at 2.2493 $\mu$m is blended with a CN feature. Lines of
Si\,{\sc i}, S\,{\sc i}, and Fe\,{\sc i} are especially prominent in the
spectrum of HD 148839 and distinctly broadened in Y Mus and UW Cen (where they
are weakly present), and absent from V854 Cen. The atomic lines seen in the RCB
stars Y Mus and UW Cen are broader than in  other stars in our sample.

Identification of  atomic lines  in the additional regions observed is a
difficult task because of the much higher density of CO and CN molecular lines.
The Na\,{\sc i} lines at 2.3348 $\mu$m and 2.3379 $\mu$m are clearly seen in all
HdC stars. Additional atomic lines are identifiable in the hotter HdC stars such
as HD 173409 (6100 K) and HD 148839 (6500 K), and  in  HD 182040 (5600 K), as
clearly seen in our Figures 3 and 4, where we have ordered the observed spectra
(from top to bottom) according to  increasing  effective temperature. The atomic
features are identified as those  lines having a similar strength in HD 182040,
HD 173409 and HD 148839. We note that  HD 173409 (6100 K) and HD 148839 (6500 K)
have almost identical spectra. By comparing with the infrared spectra of the Sun
(Wallace \& Livingston 2003) and of the giant Arcturus (Hinkle et al. 1995), we
could identify a few other atomic lines present in the spectra of the warmer HdC
stars (see Figures 3, 4 and 5). This is the case for a Mg\,{\sc i} line at
2.3328 $\mu$m, a Fe\,{\sc i} line at 2.3566 $\mu$m and a blend of Si\,{\sc i}
and Sc\,{\sc i} at 2.3579 $\mu$m.

Two other atomic lines are clearly seen at 2.3438 and 2.3443 $\mu$m (Figure 4).
These features are not seen in the Sun and Arcturus. According to the Kurucz
and  VALD-2 databases, the 2.3443 $\mu$m feature is a C\,{\sc i} line; the line
is a close doublet with components at 2.34430 $\mu$m and 2.34425 $\mu$m. Another
(single) C\,{\sc i} line is predicted at 2.3437 $\mu$m, perhaps slightly offset
from the atomic feature at 2.3438 $\mu$m: a difference of about 0.18 cm$^{-1}$
but the energy levels of upper and lower terms are apparently known to a higher
precision. These C\,{\sc i} lines are predictions from the  known energy levels
(Johansson 1966; Moore 1993)\footnote{Curiously, the NIST list of C\,{\sc i}
energy levels omits the upper levels giving the 2.34430 $\mu$m doublet but
includes the upper level giving the 2.3437 $\mu$m line.}. Wallace \& Hinkle
(2007) have recently identified a large number of C\,{\sc i} lines in the
near-IR spectrum of the Sun, showing that additional analysis using modern data
can significantly extend identification of atomic lines. Unfortunately, they did
not identify solar C\,{\sc i} lines in our regions. In any case, C\,{\sc i}
lines are expected to be stronger in our H-deficient and C-rich stars compared
to the Sun. Thus, we propose that the feature at 2.3438 $\mu$m corresponds also
to a C\,{\sc i} line (although it may be slightly blended with a Si\,{\sc i}
line). As we will see in Section 3.3, this identification is supported by our
spectrum synthesis in the warmer HdC stars.

The 2.326--2.336 $\mu$m  and 2.336--2.346 $\mu$m windows shown in Figures 3 and
4, respectively, provide a selection of CO lines. The region is crossed by P and
R lines from the $^{12}$C$^{16}$O 2-0 band with its head at 2.292 $\mu$m and the
3-1 band with its head at 2.322 $\mu$m; the strong lines are from the 3-1 band.
The 2-0 bandhead for $^{13}$C$^{16}$O occurs at 2.344 $\mu$m but is not an
obvious feature in Figure 4 for HD 137613 and HD 175893, the two HdC stars with
strong CO lines. The 2-0 bandhead for $^{12}$C$^{17}$O is at 2.3219 $\mu$m and
this isomer is a potential additional contributor to the regions shown in
Figures 3 and 4. The isomer $^{12}$C$^{18}$O is not a contributor to these
regions.

The $^{12}$C$^{18}$O 2-0 band with its head at 2.349 $\mu$m is captured  on
Figures 5 and 6 along with the $^{12}$C$^{16}$O 4-2 bandhead. There are also
$^{12}$C$^{14}$N lines in these regions. The illustrated spectra may also
include  lines from the 2-0 $^{13}$C$^{16}$O band and the 2-0 and 3-1
$^{12}$C$^{17}$O bands (the 3-1 head is at 2.351 $\mu$m). The prominence of
$^{12}$C$^{18}$O lines in the spectrum of the HdC star HD 137613  provides the
anticipated confirmation of Clayton et al.'s (2005) discovery of a high $^{18}$O
content in this star from medium resolution spectra. Similarly, our spectra
confirm Clayton et al.'s (2007)  report of the high $^{18}$O abundance for HD
175893. Indeed,  HD 137613 and HD 175893 are spectroscopic twins. Our spectrum
of S Aps (Figure 6) confirms that $^{18}$O is present (Clayton et al. 2007) but
the isotopic ratio $^{18}$O/$^{16}$O is by inspection less than in the HdC stars
HD 137613 and HD 175893.

The region shown in Figure 7 is longward of the 2-0 $^{12}$C$^{18}$O bandhead
and, thus, is a mix of $^{12}$C$^{16}$O, $^{12}$C$^{18}$O, and  $^{12}$C$^{14}$N
lines. Synthetic spectra computed for HD 137613 for the CO isomers and for  CN
are shown at the top of the figure.

\section{Abundance analysis}

\subsection{Synthetic spectra}

We have performed an abundance analysis by using spectral synthesis techniques
and extensive linelists. For this, we have  constructed H-deficient MARCS model
atmospheres for the spectroscopic effective temperatures (see Table 1) of the
stars in our sample. The MARCS models follow the prescription discussed by
Asplund et al. (1997). The input chemical composition was taken as
representative of the RCB and EHe stars, as determined by Rao \& Lambert (1994),
i.e.,  log$\epsilon$(H)=7.5 and a C/He ratio of 1\% by number
density\footnote{The abundances are normalized to log $\sum \mu_i$
$\varepsilon_{i}$ =12.15 where $\mu_i$ is the mean atomic weight of element $i$,
i.e. log$\varepsilon$(He)=11.52 and log$\varepsilon$(C)=9.52, see Asplund et al.
(1997).}. The input abundances of O and heavier elements are roughly solar. For
the HdC star HD 148839 we have used a less H-deficient model with
log$\epsilon$(H)=9.82, as suggested by Warner (1967). The input abundances of
`metals' such as Mg, Si, Ca, and Fe etc. do not greatly affect our results
because they contribute very little opacity and the more abundant C is the
primary electron donor. A small set of models were computed for C/He ratios of
0.1\% and 10\%.

For HdC stars, C/He is, in principle, measureable from infrared C\,{\sc i} and
C$_2$ lines because  He$^-$ is the dominant opacity source  with C as the
leading electron donor. Unfortunately, our spectra provide minimal
representation of these lines.  A couple of C\,{\sc i} lines contribute to our
spectra but there are open questions about their identification, especially in
the coolest stars, and the accuracy of the $gf$-values. The useful C$_2$ lines
are limited to two lines.

For RCB stars, the C/He is an ill-determined quantity from observations (Asplund
et al. 2000); the predicted equivalent width of an optical C\,{\sc i} line is
insensitive to the C/He ratio because the dominant source of continuous opacity
is photoionization of neutral carbon atoms from bound levels with an excitation
potential similar to that of the C\,{\sc i} lines. The prediction is verified in
part in that the equivalent width of a given C\,{\sc i} line varies little from
star-to-star although the equivalent width of other lines (e.g., Fe\,{\sc i} and
Fe\,{\sc ii} lines) may vary considerably. However, analysis of RCB spectra
encountered what Gustaffson \& Asplund (1996) called the `carbon problem'. The
carbon lines are weaker than predicted by the model atmosphere, i.e. ``the
product of the $gf$-value and carbon abundance provided by the C\,{\sc i} lines
is about 0.6 dex less than the product of the $gf$-value provided by quantum
chemistry and the  abundance assumed in the construction of the model
atmosphere". The `carbon problem' for RCB stars is discussed in detail by
Asplund et al. (2000). The C/He ratio is obtainable for EHe stars  for which the
carbon problem is not an issue and which provide  the  mean value of C/He =
0.0066 with a small star-to-star spread (Pandey et al. 2006). By association,
this result suggests that a C/He of 1\% may be a fair value for RCB and HdC
stars.

In addition to the input composition,  effective temperature and surface gravity
must be chosen. For the HdC stars (with the exception of HD 148839) and the RCB
stars S Aps, UW Cen and Y Mus, we adopted the T$_{\rm eff}$'s estimated from
optical spectra and reported by Asplund et al. (2000). For the HdC star HD
148839 we assumed a spectroscopic T$_{\rm eff}$=6500 K (Lawson et al. 1990). The
T$_{\rm eff}$ of 6750 K derived  from optical spectra (Asplund et al. 1998) was
chosen for the RCB star V854 Cen. The chosen T$_{\rm eff}$'s for the HdC stars
and the RCB star S Aps are those reported by Clayton et al. (2007) for  analysis
of their medium resolution near infrared spectra (their Table 2). With regard to
the gravities, a surface gravity of $\log$g$=+0.5$ in cgs units was chosen for
the HdC stars and S Aps. The RCB and EHe stars form a sequence in the ($T_{\rm
eff},\log$g) plane with $\log$g $\simeq +0.5$ expected for the HdC stars. The
surface gravity of the other three RCB stars was also assumed to be log$g=+0.5$.
Asplund et al. (2000) reported log$g=+0.75$ and $+$1.0 for Y Mus and UW Cen,
respectively. A microturbulence of 5 km s$^{-1}$ was assumed for the model
atmosphere construction. The $T_{\rm eff}$ adopted for the model atmospheres are
listed in Table 2, and, as noted above, all models have $\log$g$=+0.5$.

Synthetic spectra were generated with the TURBOSPECTRUM package (Alvarez \& Plez
1998) which shares much of its input data and routines with MARCS. In
particular, the opacity routines are consistent with those used in the
computation of the H-deficient models. The synthetic spectra were calculated for
different microturbulent velocities ($\xi$ $\sim$ 3$-$15 km s$^{-1}$ in steps of
1$-$2 km s$^{-1}$) and were convolved with a Gaussian profile with a width of 6
km s$^{-1}$, which represents the instrumental profile at the resolution of our
observations. We obtained an estimate of the microturbulence by comparing
synthetic spectra with the observed line widths of the (both weak and strong)
$^{12}$C$^{16}$O lines. For the HdC stars and the RCB star S Aps we found
microturbulent velocities of $\sim$7 km s$^{-1}$, a value in good agreement with
the mean value of 6.6 km s$^{-1}$ obtained for the cooler RCB stars in their
sample by Asplund et al. (2000) and with $\xi$=6.8 km s$^{-1}$ for HD 137613
derived by Kipper (2002) from an optical spectrum. For the hotter RCB stars Y
Mus and UW Cen, which do not show detectable CO lines, the microturbulent
velocities of 10 and 12 km s$^{-1}$ (Asplund et al. 2000), respectively, were
adopted. These higher microturbulences are consistent with the broader
appearance of the atomic lines in the 2.251 $\mu$m region. In addition, we need
to convolve the synthetic spectra for these two RCB stars with a larger Gaussian
profile ($\sim$20$-$25 km s$^{-1}$) in order to take into account a higher
macroturbulence. We will adopt a maximum error of $\pm$2 km s$^{-1}$ for the
microturbulence in computing our derived chemical abundances (see Table 3).

Synthetic spectra were computed using a list of atomic and molecular lines. Key
contributors of lines are the CO, CN and C$_2$ molecules. Line positions for
$^{12}$C$^{16}$O, $^{12}$C$^{17}$O, $^{12}$C$^{18}$O, and $^{13}$C$^{16}$O were
computed from molecular constants for $^{12}$C$^{16}$O (Huber \& Herzberg 1979)
with constants for isomers computed via standard relations. Oscillator strengths
were taken from Goorvitch (1994). The CO dissociation energy was put at
D$_{0}$(CO)=11.09 eV for all isomers. A line list for  $^{12}$C$^{14}$N and its
isomers  was kindly provided by B. Plez (private communication; see Hill et al.
2002 for a discussion of Plez's linelist). The dissociation was put at
D$_{0}$(CN)=7.76 eV (Costes et al. 1990). Lines of the $\Delta v$ $=-2$ sequence
of the C$_2$ Phillips system cross observed spectral windows and, as noted in
Section 2, two or three are present unblended or almost so in the spectra of
several of our stars. Line positions are taken from Davis et al. (1988) and
oscillator strengths calculated from band oscillator strengths (Kokkin et al.
2007) and  H\"{o}nl-London factors (Lambert et al. 1986). The empirical
dissociation energy of $D_0 = 6.297$ eV measured by Urdahl et al. (1991) is
used.

For atomic lines, the primary source of information was the VALD-2 database
(Kupka et al. 1999). As noted above,  identification of atomic features was made
using the infrared atlas (R$\sim$100,000) of the giant Arcturus (Hinkle et al.
1995) and the infrared solar spectrum (Wallace \& Livingston 2003). The
$gf$-value of an atomic lines was adjusted so that the predicted line strength
matched the observed strength in the Arcturus atlas. For this, a MARCS model
atmosphere was used with Arcturus's fundamental parameters determined by Decin
et al. (2003): T$_{\rm eff}$=4300 K, $\log g $=1.50, [M/H]=$-$0.5,  $\xi$=1.7 km
s$^{-1}$, C/N/O=7.96/7.55/8.67, and $^{12}$C/$^{13}$C=7. These parameters are in
good agreement with other determinations reported in the literature. Finally, we
also test that  our molecular linelists reproduce well the observed
high-resolution near-IR spectrum of Arcturus. The agreement is especially good
for CO where the $gf$-values and wavelengths are well known. The agreement is
also good for the few CN features around 2.251 $\mu$m observed in the Arcturus's
spectrum and this region will be used for our N abundance determination. We note
that our CN linelist contains a few incorrect wavelengths  but, in general, 
affected lines are easily identified from a comparison of synthetic and observed
spectra. In addition, when stellar CN is strong there could be  more stellar CN
lines than those present in our linelists (see Section 3.2.3). This is the case
of the 4-6 CN Red System band which has recently been shown to have a lot of
lines in our regions (Davis et al. 2005). Unfortunately, the $gf$-values of
these CN lines are unknown. The C$_2$ lines are absent, as expected, from the
Arcturus spectrum.

Synthetic spectra were computed and matched to the observed spectra. Our primary
goal was to determine the isotopic ratios for C, N, and O, as the determination
of these ratios is almost independent of the adopted model parameters. The
secondary goals were to constrain the C, N, and O elemental abundances and to
determine the abundances of heavier elements, principally Na and S.

\subsection{The isotopic ratios for C, N, and O}

For the HdCs HD 137613, HD 175893, and the RCB S Aps, we are able to obtain
reliable estimates of the isotopic ratio $^{16}$O/$^{18}$O. For the HdC star HD
182040, we estimate the    $^{16}$O/$^{18}$O ratio from a few weak features. For
HdC stars HD 148839 and  HD 173409 and the other RCBs, identification of CO in
any isotopic combination is  uncertain so that we are decline to give isotopic
ratios. For three HdCs and S Aps, lower limits are determined for the ratios
$^{12}$C/$^{13}$C, $^{14}$N/$^{15}$N, and $^{16}$O/$^{17}$O.

\subsubsection{HD 175893}

In Figures 8 to 12, we illustrate the fit of synthetic spectra to the observed
spectra for HD 175893. The 2.246--2.256 $\mu$m region (Figure 8), as noted
above, is free of first-overtone CO lines and rich in CN Red System lines. The
computed spectrum (red line) with $\log \epsilon$($^{14}$N) = 9.2,  the input
$^{12}$C abundance ($=9.52$), and the total O abundance ($=8.7$) from the
isotopes $^{16}$O and $^{18}$O is a good fit to the observed spectrum. A  few
lines in the observed spectrum are absent from the adopted linelist. For a few
CN lines, a small revision of their adopted wavelengths is indicated but not
here applied. The synthetic spectrum (blue line) corresponding to a reduction of
the N abundance to 8.7 with adopted C and O abundances left unchanged reveals
which lines  arise wholly or primarily from $^{12}$C$^{14}$N. The fact that the
majority   of the line depths of many  lines vary approximately in the ratio of
the change of the N abundance shows that the lines are not severely saturated.
The  N abundance derived from CN lines is sensitive to the adopted C and O
abundance. The CN number density in the atmosphere depends on the product of the
partial pressures of C and N. The partial pressure of C is determined by both
the C and O abundances  through the formation of CO. The partial pressure of N
is determined by the abundance of N with minor depletion of free nitrogen from
N$_2$ formation and is insensitive to the C and O abundances. This region is
used to set the limit $^{14}$N/$^{15}$N $>10$. This lax limit on $^{15}$N arises
because the strongest CN lines are just 30 per cent deep and the line density of
$^{12}$C$^{14}$N lines is high.

The $^{16}$O abundance is determined from the 2.326--2.336 $\mu$m and
2.336--2.346 $\mu$m regions shortward of the $^{12}$C$^{18}$O 2-0 bandhead.
Synthetic and observed spectra for these regions for HD 175893 are shown in
Figures 9 and 10. The syntheses adopt the $^{14}$N abundance (= 9.2) from the
fit to the CN lines in  Figure 8, and the input C abundance (=9.52). Syntheses
are shown for the same total O abundance (the sum of $^{16}$O and $^{18}$O
abundances) but two choices of  isotopic ratio and, hence, two different
$^{16}$O abundances. The $^{12}$C$^{16}$O lines are well fit for an abundance
$\log \epsilon$($^{16}$O)=8.1 (i.e., corresponding to the isotopic ratio
$^{16}$O/$^{18}$O=0.3 and  total O abundance of $\log \varepsilon$(O)=8.7, see
below). The $^{13}$C$^{16}$O 2-0 bandhead at 2.3441 $\mu$m  is just captured on
Figure 10. A limit $^{12}$C/$^{13}$C $> 10$ is set using synthetic spectra with
different abundances of $^{13}$C.

The $^{12}$C$^{18}$O 2-0 band is captured on Figures 11 and 12. The synthetic
spectra give us a $^{16}$O/$^{18}$O ratio of 0.3: the  total O ($^{16}$O $+$
$^{18}$O) abundance is $\log \varepsilon$(O)=8.7 with  $\log
\epsilon$($^{16}$O)=8.1 and $\log \epsilon$($^{18}$O)=8.6. The best synthetic
spectra displayed in these figures are computed for the input C abundance and
the CN lines are included with the N abundance derived from Figure 8. The
sensitivity of the derived $^{16}$O/$^{18}$O ratio to the adopted atmospheric
parameters is summarized in Table 3. Our measurement of this isotopic ratio is
in good agreement with Clayton et al.'s (2007) estimate of $^{16}$O/$^{18}$O=0.2
from the intensities of the two bandheads, as measured off medium resolution
spectra.

Our estimate of the $^{17}$O abundance is obtained from the $^{12}$C$^{17}$O 3-1
band with its head at 2.3511 $\mu$m and a few 2-0 and 3-1 $^{12}$C$^{17}$O lines
longward of the head. By spectrum synthesis (Figure 11), we find no convincing
evidence for the presence of $^{12}$C$^{17}$O and set the limit
$^{16}$O/$^{17}$O $> 50$, the first such estimate for a HdC star.

\subsubsection{HD 137613}

Analysis of the spectra followed the procedure just described for HD 175893. For
the input C abundance (=9.52), the N abundance from the CN lines is found to be
$\log\epsilon$(N)=9.4 and the total O abundance to be $\log\epsilon$(O)=8.7.
Selected comparisons of synthetic and observed spectra are provided in Figures
13, 14 and 15. With a ratio $^{16}$O/$^{18}$O=0.5, HD 137613 is slightly less
$^{18}$O-rich than HD 175893, a result consistent with Clayton et al.'s (2007)
finding too. Our limit on the $^{17}$O abundance is the same as for HD 175893.
Our limit on $^{12}$C/$^{13}$C ($>10$) is not competitive with determinations
from CN Red System lines near 8000 \AA: Kipper (2002) reported that `the
$^{12}$C/$^{13}$C ratio turns out be at least 90' and Fujita \& Tsuji (1977)
boldly set the ratio at 500 or greater. An absence of  lines from
$^{13}$C-containing molecules in optical spectra was noted by Warner (1967) for
all five HdC stars.

\subsubsection{HD 182040}

Examination of Figure 1 shows that HD 182040 has CN lines of similar strength to
those displayed by the twins HD 137613 and HD 175893. As   mentioned in Section
3.1, our CN linelist is not complete. This incompleteness (wavelength errors,
missing lines) is evident for HD 182040 where the CN molecule dominates the
spectrum but  an excellent  fit to the observed spectrum cannot be obtained.
Synthesis of the 2.251 $\mu$m region (Figure 16) provides a satisfactory fit to
the observed spectrum when the N abundance is $\log\epsilon$($^{14}$N)=9.2 with
the input C abundance (=9.52) and the total O abundance set by the CO lines at
$\log\epsilon$(O)=8.0. Again, we set the limit $^{14}$N/$^{15}$N $>10$.

Inspection of Figure 3, 4, and 7 shows that weak $^{12}$C$^{16}$O lines are
present. Many other $^{12}$C$^{16}$O are obviously too strong;  CN is the
probable contaminant. In Figure 7, inspection similarly  shows the presence of
unblended $^{12}$C$^{18}$O lines with several others strengthened by blends.
Unfortunately, the 2-0 $^{12}$C$^{18}$O bandhead was not completely captured on
our spectra. We base our $^{16}$O/$^{18}$O determination on the cleanest
$^{12}$C$^{16}$O and  $^{12}$C$^{18}$O lines (Figure 17). Anticipation of an
distinctive 2-0 $^{12}$C$^{18}$O is strongly suggested by Figure 5. The ratio
$^{16}$O/$^{18}$O=0.5 is obtained, a value consistent with Clayton et al.'s
(2007) estimate of 0.3.

Our limit on the $^{12}$C/$^{13}$C ratio ($>10$) is consistent with the more
stringent limit $^{12}$C/$^{13}$C $>100$ provided by Fujita \& Tsuji (1977). Our
lower limit to the $^{16}$O/$^{17}$O ratio is 8.

\subsubsection{HD 173409}

Unfortunately, this star was not observed in the 2.251 $\mu$m region and we
could not obtain an estimate of the N abundance from a region free of CO lines.
However, examination of our spectra (Figures 3 and 4) shows that weak
$^{12}$C$^{14}$N  are clearly present in this star with T$_{\rm eff}$=6100 K.
Examination of the same figures shows that $^{12}$C$^{16}$O lines are weak, if
not absent. A similar conclusion applies from Figure 7 to $^{12}$C$^{18}$O
lines. Our spectra of HD 173409 did not capture the 2-0 $^{12}$C$^{18}$O
bandhead at 2.349 $\mu$m and at longer wavelengths the mix of $^{12}$C$^{16}$O,
$^{12}$C$^{18}$O and $^{12}$C$^{14}$N lines hampers identification of the CO
isomers.

We have used the unblended $^{12}$C$^{14}$N and possible $^{12}$C$^{16}$O
features in the region 2.326$-$2.346 $\mu$m to estimate the N/O ratio (for the
assumed C abundance). For this, we fixed the total O abundance to $\log
\varepsilon$(O)=8.7 (as obtained for the HdC stars HD 137613 and HD 175893) and
we calculated the $^{14}$N abundance needed to fit the cleanest $^{12}$C$^{14}$N
lines.  We selected a few $^{12}$C$^{14}$N lines (e.g. those around 2.342
$\mu$m; see Figure 18) for which a perfect agreement with the N abundance
derived from the 2.251 $\mu$m region was obtained in the other HdC stars. By
this procedure, the  N abundance from the CN lines is found to be
$\log\epsilon$(N)=9.2, giving a N/O ratio of $\log\epsilon$(N/O)=$+$0.5. Through
a family of syntheses,  we conclude that determination of the  $^{16}$O/$^{18}$O
ratio is very difficult, if not impossible, to determine from the present
observations.

\subsubsection{HD 148839}

In spite of the higher effective temperature (6500 K) for this star, the CN
lines are of similar strength to those of HD 173409 (6100 K). The strongest
$^{12}$C$^{14}$N features are clearly seen in the 2.251 $\mu$m spectrum (Figures
1 and 2). Our spectrum synthesis in this region (Figure 19) gives a high N
abundance of 9.8 with the input C abundance (=9.52) and a total O abundance of
9.2, the best estimate from tentative identifications of CO lines.

Inspection of Figures 3 and 4 suggest that $^{12}$C$^{16}$O lines are weak but
consistently present. Evidence for $^{12}$C$^{18}$O lines is less certain. One
may contrast the two stars HD 148839 and HD 182040: $^{12}$C$^{16}$O lines
appear present in HD 148839 and weaker in HD 182040 with the opposite being the
case for the $^{12}$C$^{18}$O lines. The lower limit for the ratio
$^{16}$O/$^{18}$O is probably 2. This is intriguing  because  HD 148839 is
Li-rich (Rao \& Lambert 1996) and synthesis of $^{18}$O resulting from the DD
merger  would seem difficult to reconcile with synthesis of Li (see below).

\subsubsection{S Aps}

The CN lines are of similar strength to those in HD 137613 and HD 175893. The
synthetic spectrum fitted to CN lines in the 2.251 $\mu$m region (Figure 21)
gives a N abundance of 9.6 with the input C abundance (=9.52) and the total O
abundance (=9.4) from the CO lines. Synthesis of the 2.354 $\mu$m region gives a
good overall fit to the observed spectrum (Figure 22). The lower limits on the
$^{14}$N/$^{15}$N ($>$10), and $^{16}$O/$^{17}$O ($>$80) ratios are comparable
to those set for HD 137613 and HD 175893. We could not set a lower limit on the
$^{12}$C/$^{13}$C ratio because this star was not observed around the 2-0
$^{13}$C$^{16}$O bandhead at 2.344 $\mu$m.

Inspection of Figures 6 and 22 shows that the $^{16}$O/$^{18}$O ratio for S Aps
is greater than unity in contrast to the values of less than unity prevailing in
HD 137613 and HD 175893. The derived isotopic ratio is $^{16}$O/$^{18}$O=16, a
value larger than Clayton et al.'s (2007) estimate of 4. Our result is somewhat
dependent on the adopted microturbulence given that the available
$^{12}$C$^{18}$O lines are systematically weaker than the $^{12}$C$^{16}$O
lines. S Aps is the only RCB star in our sample for which the isotopic O
composition could be measured. This is unfortunate because S Aps possibly offers
the hint that the RCBs may be less enriched in $^{18}$O than the HdCs.

\subsection{Elemental abundances - C, N, and O}

For the analysis of the CN and CO lines, we assumed that the C abundance is
equal to the input C abundance of the adopted model atmosphere, which for
models with C/He=1\% is a C abundance of 9.52.  As long as the assumed C
abundance is greater than the total O ($^{16}$O+$^{18}$O), the CO number density
is not very dependent on the C abundance. In the case of CN, the number density
is dependent on the difference between the C and O abundances thanks to the
dominant role of CO formation, and on the N abundance with N$_2$ playing the
controlling role in setting the partial pressure of N. The abundance of C (i.e.,
the C/He ratio) plays a secondary role in setting the continuous opacity.

A check on the input C abundance is possible from the C\,{\sc i} lines. Both
C\,{\sc i} lines are usable in the two hottest stars HD 173409 and HD 148839.
Unfortunately, the C\,{\sc i} line at 2.3443 $\mu$m is strongly blended with
molecular features in the cooler HdC stars although the other C\,{\sc i} line at
2.3438 $\mu$m, if correctly identified, seems to be unblended. The wavelength
interval providing the C\,{\sc i} lines was not observed for S Aps and the other
RCB stars.

Analysis of the five HdC stars using models computed for C/He=1\%, i.e., an
assumed abundance $\log\epsilon$(C)=9.52, provide from the C\,{\sc i} lines
carbon abundances that average only $+0.1$ dex greater than the assumed
abundance with a range from $-0.3$ to $+0.4$ dex. The extremes of the range are
provided by HD 137613 at $-0.3$ dex and HD 182040 at $+0.4$ dex. This is a
satisfactory level of consistency between the C abundance adopted in the models'
construction and that provided from the C\,{\sc i} line(s). Thus, the C\,{\sc i}
lines do not exhibit the carbon problem highlighted by Asplund et al. (2000) in
their analysis of the optical spectra of RCBs. The C abundance obtained depends
on the C/He ratio adopted for the models. At C/He=10\%, the C abundance from
C\,{\sc i} lines is about 0.2 dex larger. For C/He$<1$\% models, there is a
point at which the  input C abundance is less than the O abundance needed to
account for the CO lines, that is the stars are O-rich which is in stark
contrast to the appearance in the optical and infrared spectra of strong bands
of C-containing molecules. For the coolest stars ($T_{\rm eff} < 5500$ K), C/He
ratios of less than about 0.5\% result in an input C abundance less than the
derived total O abundance. For the mean C/He ratio ($=0.7$\%) of the EHes, the
stars will be C-rich but barely so.

The C$_2$ Phillips system lines 0-2 Q(34), 1-3 Q(8) and 1-3 Q(10) provide
another opportunity to obtain a C abundance. For the three HdC stars with
detectable  C$_2$ lines, the C$_2$ based abundance is 0.8 (HD 137613), 0.7 (HD
175813), and 1.0 (HD 182040) dex larger than the input abundance. This is `a
carbon problem'. This particular carbon problem appears to be largely resolved
for models with C/He$=10$\%; the input abundance is increased by 1.0 dex over
that of the C/He$=1$\% models but the C abundance needed to fit the C$_2$ lines
is only slightly increased over that from the C/He=1\% models.

Not only is a C/He=10\% ratio at odds with the much lower ratio of the EHes that
is supposed to be generally applicable to HdC stars but the C abundance from the
C\,{\sc i} lines which matched the input abundance of the C/He=1\% models is
only slightly increased by use of the C/He=10\% models and, therefore, removal
of the carbon problem posed by the C$_2$ lines is accompanied by the creation of
a carbon problem of about 1 dex from the C\,{\sc i} lines.

Here, our emphasis on the $^{16}$O/$^{18}$O ratio plausibly  allows us to
postpone a search for the solution to these carbon problems. Isotopic ratios are
insensitive to the adopted C/He ratio and may not depend greatly on the correct
solution to the carbon problems. Certainly, the presence of a very low
$^{16}$O/$^{18}$O will not be denied.

In our analysis, we, as noted above, adopt a model's input C abundance in
extracting the N and O abundances from the CN and CO lines. Extracted abundances
change by less than 0.2 dex when models with C/He ratios from 0.1\% to 10\% are
used for the analysis. This is certainly a fortunate circumstance. The derived
total elemental C, N, and O abundances given in  Table 2  are for C/He=1\% and
the assumed input composition of heavier elements (i.e.
log$\varepsilon$(Fe)=7.2, log$\varepsilon$(Na)=6.8, log$\varepsilon$(Si)=7.7,
etc., Asplund et al. 1997). The formal error in the total N and O abundances
taking into account variations of the atmospheric parameters used in the
modeling are estimated to be of the order of 0.3$-$0.4 dex.

The HdC HD 137613 was analysed previously by Kipper (2002) with C, N, and O from
C$_2$ Swan system, CN Red system, and O\,{\sc i} lines. Our derived abundances
for HD 137613 are in poor agreement with those reported by Kipper (2002) from an
abundance analysis of the 4780 \AA\ to 6400 \AA\ spectrum. For ($\log
\epsilon$(C),$\log \epsilon$(N), $\log \epsilon$(O)), our results are
(9.5,9.4,8.7) whereas Kipper reported values of (9.3,8.7,7.6). Kipper (2002)
noted  that his analysis of C$_2$ Swan system bands led to a C abundance 0.2 dex
less than the input abundance, which considering the uncertainties in the
analysis indicates that the carbon problem is diminished or absent for this HdC.
(Kipper, however, considered that the carbon problem was present.) Detailed
discussion of the abundance differences between this and Kipper's study is not
attempted here where we focus on the isotopic O ratios.

Asplund et al. (2000) obtained C, N, and O abundances for Y Mus in their
analysis of warm RCBs. Their C abundance exhibits the carbon problem in that the
optical C\,{\sc i} lines gave $\log\epsilon$(C) = 8.9 for models with the input
abundance of 9.5. Their N abundance of 8.8 contrasts with our value of 9.4.
Unfortunately, we are unable to determine the O abundance.

\subsection{Na and S abundances}

Several atomic lines (Na\,{\sc i}, S\,{\sc i}, Mg\,{\sc i}, Fe\,{\sc i}, C\,{\sc
i}, etc.) are identified in our spectra of the warmer HdC stars. Na and S are
the only elements showing two (or more) unblended absorption lines for the five
HdC stars and their abundances will be more reliable. Abundances of Na and S in
Table 2 are from models with C/He=1\%. These values are unchanged if the
metallicity (abundance of elements heavier than O) is decreased from its
quasi-solar value; the electrons contributing to the  He$^-$ continuous opacity
are donated by C atoms primarily. Obviously with He$^-$ as an opacity source,
the derived Na and S abundances are dependent on the C/He ratio assumed in
constructing the model atmosphere: the change from C/He=1\% to 10\% results in
about a 0.6 dex increase in the Na and S abundances.

Sodium abundances were obtained for the five HdC stars. The region providing the
2.33 $\mu$m Na\,{\sc i} doublet was not observed for the  RCBs. The Na
abundances from the C/He=1\% models run from $\log\epsilon$(Na)= 6.5 to 7.0 for
a mean abundance of 6.8. The S abundance has been estimated for all of our stars
except HD 173409 for which the region providing the S\,{\sc i} lines was not
observed. The mean S abundance is $\log\varepsilon$(S)=7.5 for the HdC stars and
6.8 for the RCB stars S Aps and Y Mus. The photospheric spectrum of UW Cen is
diluted by circumstellar emission (see above) and, hence, the derived S
abundance is an underestimate.

For the HdCs, the Na and S abundances are supra-solar from the C/He=1\% models;
if we assume [Na/Fe] = [S/Fe] = 0, the implied Fe abundances  are $+0.2$ to
$+0.5$ dex above solar. The Mg\,{\sc i}, Si\,{\sc i}, and Fe\,{\sc i} generally
confirm these Fe overabundances. However, the RCBs S Aps and Y Mus give the
sub-solar S abundance expected of EHes and RCBs. The Fe abundance for Y Mus from
one Fe\,{\sc i} line is also sub-solar. The S abundance in Table 2 for Y Mus is
within 0.1 dex of the value reported by Asplund et al. (2000) from optical
spectra, a difference well within the errors of measurement.

Few direct comparisons with the literature are possible for these Na and S
abundances. The Na abundance for HD 137613 is 0.9 dex greater than Kipper's
(2002) from optical spectra. Kipper did not determine a S abundance but his
abundances for other $\alpha$-elements (Mg, Si, Ca, and Ti) are $-$0.8 dex less
than solar while our S abundance is slightly supra-solar.

For the C/He=10\% models, the Fe range  for the HdC stars is lifted to $+0.8$ to
$+1.1$ above solar, an impossible interval! Models with a C/He below 1\% will
result in lower Na and S abundances but, as noted above, there is a lower limit
to the C/He ratio not too much less than 1\% below which the HdC appears oxygen
rich. That limit will not suffice to reduce Na and S below their solar
abundances. In short, the Na and S abundances present an interesting problem
within the constraints set by the assumption of local thermodynamic equilibrium
for the models and the analysis of the lines.

\subsection{Compositions along the HdC, RCB, and EHe sequence}

If the  working hypothesis that the three classes of H-deficient luminous stars
share a common evolutionary scenario, their chemical compositions should also
share some common characteristics. Since these stars are supergiants with
presumably deep convective envelopes that may achieve dredge-up of nuclear
processed material from a He-burning shell, there may be changes of composition
along the evolutionary track(s) linking the HdC, RCB, and EHe stars.

It is evident that EHe stars  are not all of the same composition. The
metallicity spans over two orders of magnitude with relative abundances of the
elements which are unaffected by changes during stellar evolution  having the
values shown by normal (H-rich) stars (Pandey et al. 2006). Elements affected by
evolution include He, C, N, O and, in a few cases, the $s$-process elements. The
few EHes and RCBs having extraordinarily high Si/Fe and S/Fe ratios are termed
`minority' EHes and RCBs (Rao \& Lambert 1994) are ignored here. HdC stars all
appear to be majority representatives.

The EHe stars are a useful reference because they are immune to the carbon
problem affecting the warm RCB stars. Abundance information on EHes is taken
from Pandey et al. (2001), Pandey et al. (2006), and Pandey \& Reddy (2006). As
shown by Pandey et al. (2006), addition of data drawn from the literature for
other EHes would not affect our  conclusions. In Figure 23, N abundances for EHe
stars are shown as a function of the Fe abundances.   The EHes form a sequence
in this figure with, as shown by Pandey et al. (2006), the N abundance being
essentially equal to the sum of the initial C, N, and O abundances (the solid
line in the figure) expected from the Fe abundance which is assumed to be
unaffected by the ravages of stellar evolution. The interpretation is that the
nitrogen is contributed by a layer in which CNO-cycling has converted initial
CNO to N and that this layer is now the major component of the HdC
atmosphere/envelope. The HdCs placed on this figure using the Fe abundance
inferred from the mean of their Na and S abundances lie on the extension of the
EHes sequence; they are  apparently remarkably Fe-rich and N-rich. On the
assumption that the initial CNO abundances scale linearly with the initial Fe
abundance for supra-solar metal-rich compositions, the HdCs are also consistent
with the above idea about the origin of N. The warm RCB stars (not shown in
Figure 23) show a similar trend for N with Fe with data from Asplund et al.
(2000), and Rao \& Lambert (2002, 2008). Three minority RCBs anchor the trend at
the low Fe end. The RCB N-Fe relation is somewhat offset from the EHe trend:
superposition requires either a shift of the RCB N abundances by about $-0.4$
dex, or the Fe abundances by $+0.4$ dex, or a mix of N and Fe shifts. The shifts
may reflect systematic errors associated with the abundance analysis that left
the RCB's carbon problem unresolved.

In Figure 24, the O abundances of EHe stars are shown as a function of the Fe
abundances. The O abundances show a considerable scatter at all well-sampled
values of the Fe abundance, a scatter not shown by the N abundances. The range
in O abundances is not dissimilar for the HdC and EHe stars. The isotopic mix of
the O in EHes  is unknown. For three of the HdC stars, we know that $^{18}$O not
$^{16}$O is the dominant isotope. Anticipating that $^{18}$O is synthesized from
$^{14}$N (see next section), it is important to note that the ratio of
$^{18}$O/N varies from star to star. This ratio for the four stars with
detections of C$^{18}$O ranges from a high of 0.25 to a  low of 0.04.
 
It is apparent from Figure 24 that the  majority of the O abundances equal or
exceed the  initial O abundance predicted by the Fe abundance and the usual O-Fe
relation for initial abundances. (The warm RCBs -- not shown in Figure 24  --
also show a scatter in the O versus Fe diagram and, as for the EHes, the
isotopic mix of the O in unknown. On average, the RCB stars are offset from the
EHe stars by about $+$0.6 dex in O.) The O abundances of the EHe and HdC stars
exceed the low abundances expected for $^{16}$O after CNO-cycling has converted
initial CNO to N. Survival of $^{16}$O requires a temperature of less than about
15 million degrees. The $^{16}$O after CNO-cycle equilibrium is depleted by a
factor of about 0.9 dex for cycle operation at 20 million degrees, and
depletion   increases with increasing temperature.  Once O depletion occurs, the
$^{16}$O/$^{18}$O ratio becomes very high ($\sim 10^4$ at 20 million degrees)
and the $^{17}$O abundance exceeds that of $^{18}$O ( by a factor of $\sim 5$ at
20 million degrees) (Arnould, Goriely, \& Jorissen 1999). Clearly, the  nitrogen
and oxygen abundances of the  EHe, RCB, and HdC stars seem unlikely to both be
residues from H-burning via the CNO-cycles.\footnote{Note that the high
($>$10) $^{12}$C/$^{13}$C ratio present in all HdC and RCB stars studied so far
also rules out the H-burning products in these stars.} This conclusion is
absolutely certain for the HdC stars where O isotopic abundances are available.

\section{Sites for the synthesis of $^{18}$O}

Discovery of a dramatic enrichment of $^{18}$O in HD 137613 by Clayton et al.
(2005) was anticipated by Warner (1967). Clayton et al.'s (2007) extension of
the discovery to two other HdC stars and four RCBs showed that  theories of the
evolution of these H-deficient stars must provide for the common occurrence of
abundant $^{18}$O. Theory in this discipline has two parts: a mechanism of
nucleosynthesis that provides  abundant $^{18}$O and a recipe for stellar
evolution that leads to a H-deficient object rich in $^{18}$O.

The likely heart of the mechanism of $^{18}$O synthesis was identified by
Warner (1967): $^{14}$N, the dominant ash of H-burning, is converted to $^{18}$O
by $^{14}$N$(\alpha,\gamma)^{18}$ F$(\beta^+\nu)^{18}$O in the run up to
He-burning but with further increases in temperature  another $\alpha$-capture
turns the $^{18}$O to $^{22}$Ne prior to ignition of He-burning. The $^{18}$O
yield per initial $^{14}$N is obviously dependent on the physical conditions
(temperature, density, and composition) and their temporal variation. If high
temperatures persist for a long period, $^{14}$N is converted completely to
$^{22}$Ne and the $^{18}$O abundance after achieving a maximum declines to very
low levels.

Although the mechanism of $^{18}$O synthesis appears broadly understood, the
stellar evolution that leads to triggering of the mechanism cannot claim to be
understood yet. In large part, the puzzle concerns the site at which $^{18}$O is
made from $^{14}$N and the source of that $^{14}$N.

Clayton et al. (2005) in their discussion of possible origins for the $^{18}$O
in HD 137613 proposed that its precursor was a star entering the AGB phase of
its life.  Such a star after He-core burning has a thin $^{18}$O-rich layer
between the C-O core and the  H-rich envelope. Then, if severe mass loss is
invoked to remove the envelope and expose the $^{18}$O-rich layer, a HD
137613-like star is produced. This invocation of severe mass loss early on the
AGB is not only unsupported theoretically by a possible mechanism but may be
supposed to produce H-poor stars with a wide range of $^{18}$O abundance in
contrast to the indication from Clayton et al. (2007) that severe  $^{18}$O
enrichment seems common among the small sample of  HdC stars and some RCBs.
Clayton et al. (2007) recognized these weaknesses of their 2005 proposal.

Aside from this identification of H-deficient stars with early-AGB stars, there
are, as noted in the Introduction,  two scenarios in play to account for H-poor
luminous stars -- the FF and DD scenarios. Clayton et al. (2007) note that the
FF scenario  is not expected to achieve an overproduction of $^{18}$O because
either the $^{14}$N is burnt to $^{22}$Ne and, in cases where a H-rich surviving
envelope of the post-AGB star is ingested by the final He-shell flash,  $^{18}$O
is efficiently destroyed by proton-capture.  This expectation is supported by
observations of Sakurai's object (V4334 Sgr), a star identified as a FF product,
showing  the low $^{12}$C/$^{13}$C ratio characteristic of H-burning. Asplund et
al. (1997) found the low ratio from analysis of C$_2$ Swan bands near 4740 \AA.
Their value was confirmed by Pavlenko et al. (2004) from high-resolution spectra
of strong first-overtone CO bands. Although these authors do not comment on the
presence of $^{12}$C$^{18}$O bands, it is evident from their illustrated
observed and synthetic spectra that the $^{12}$C$^{18}$O 2-0 band must be very
weak and the  $^{16}$O/$^{18}$O ratio must be high, i.e., Sakurai's object
differs markedly in this regard from the HdCs with strong CO bands, as Clayton
et al. (2007) remark on the basis of the high-resolution CO spectrum shown first
by Geballe et al. (2002).

Although some H-poor stars are outcomes of the FF scenario with Sakurai's object
being the most recent notable discovery, studies of the C, N, and O elemental
abundances for RCB and EHe stars favor the DD scenario. This partiality for the
DD scenario was developed without knowledge of the puzzle offered by $^{18}$O;
the stars in the sample are too warm for CO molecules to survive in their
photospheres.\footnote{Tenenbaum et al. (2005) found no correlation between
dust formation and CO band strength. However, there is the intriguing
possibility that CO molecules may be detectable in circumstellar gas at and
around the time a RCB star is in decline. Cold C$_2$ molecules have been
detected at this time for two RCBs (Rao \& Lambert 2000, 2008).}

In the DD scenario, a He white dwarf merges with a C-O white dwarf as emission
of gravitational radiation causes shrinkage of the binary's orbit. The He white
dwarf is disrupted and forms a coating on the C-O white dwarf. Conversion of
gravitational potential to heat causes this coating to swell to supergiant
dimensions. Helium-burning maintains the swelling of the star for a brief period
to provide a H-deficient luminous star, a EHe, RCB, or a HdC, before descent to
the white dwarf cooling track. There has not yet been a consistent discussion of
the evolution, internal  nucleosynthesis, and changes in surface composition
from the time before the merger through to the descent of the H-poor supergiant
on to its white dwarf cooling track -- it is a {\it difficult} theoretical
problem. Thus, `toy models' have to be explored as to whether they might sustain
high $^{18}$O abundances. Three such models are discussed here.

\subsection{Toy models}

\subsubsection{Model A}

 The simplest of the toy models  supposes that the merger is  cold (i.e., no
nucleosynthesis during the merger) and that the composition of the resultant
supergiant is determined solely by mixing the He white dwarf with the surface
layers of the C-O white dwarf, i.e., the surface composition of the supergiant
is unaffected by its deep convective envelope and dredge-up of
nuclear processed from  within the supergiant.

Saio \& Jeffery (2002) and Pandey et al. (2006) argue that this recipe with
certain assumptions about the compositions and masses of the He white dwarf and
the perturbed surface layers of the C-O white dwarf  provides a
semi-quantitative explanation for the CNO abundances of EHe and RCB stars but
without, of course, a consideration that the oxygen may not be $^{16}$O, as
assumed in the recipe,  but $^{18}$O. In acceptable recipes, principal
ingredients of the envelope of the merged star are about ten parts from the He
white dwarf and one part from the outer He-rich  layers of the C-O white dwarf.
These proportions are set by the observation that the C/He $\sim 0.01$ for the
EHe stars and the anticipated He and C abundances of the He and C-O white dwarf
contributions. Added according to taste may be minor amounts of normal
composition material from a H-rich layer atop the He-white dwarf and a small
contribution from    the outer    layers of the C-O white dwarf immediately
below the star's He-rich layer. The carbon, as $^{12}$C, in such recipes is
provided by either the He-rich skin and/or the C-O rich layers of the C-O white
dwarf below this skin. About 90\% of the nitrogen in these recipes is provided
by the He white dwarf and about  10\% by the old He-shell of the C-O white dwarf
with an abundance in both cases, it is assumed, equal to the sum of the initial
C, N, and O abundances, as a result of CNO-cycling. Thus, the observation that
the N abundance is equal to the sum of the initial C, N, and O is reproduced by
the model. Oxygen, as $^{16}$O, in these recipes is contributed  by the C-O
white dwarf from its He-rich layer or the scooped up outer layers of its C-O
rich core. With plausible ratios for the C/O abundances of these layers it is
possible to reproduce the O abundances of the EHe stars.

Given the number of the free parameters in the recipe, it is not surprising that
the observed  He, C, N, and O abundances of the EHe stars can be reproduced.
Indeed, one might say that it is the observed abundances that restrict the
choices of mass ratio of material from the He and C-O white dwarfs, and the
ratio of mass fractions of C to O in the material contributed by the C-O white
dwarf. Yet,  it is important to ask where in a family of such recipes might
$^{18}$O be included to provide a star after the cold merger with an abundance
comparable to or exceeding that of $^{16}$O?

Two suspects may be arraigned: the He white dwarf and the He-shell of the C-O
white dwarf. Perhaps, a fraction of the He white dwarf in its evolution before
onset of electron degeneracy achieved temperatures sufficient to convert
$^{14}$N to $^{18}$O. The reductions of the $^{14}$N  implied  by the   observed
$^{18}$O/N ($<0.25$) ratios do not within the observational errors negate the
proposition that the N abundance within the observational errors are equal to
the sum of the initial C, N, and O abundances. Perhaps, the more plausible site
for the $^{18}$O is the He-shell around the C-O white dwarf. In evolution to the
white dwarf stage, this layer was a bridge between H-burning on one side and
He-burning on the other side and, therefore, a part of the He-shell may have
experienced partial conversion of N to $^{18}$O. The $^{16}$O/$^{18}$O ratio
then depends on the degree of conversion of N to $^{18}$O in the He-shell and
the contributions to $^{16}$O from the He-shell and the C-O white dwarf.

\subsubsection{Model B}

In this family of models, the merger is not cold but hot, i.e., nucleosynthesis
occurs during the merger, and detailed studies of the
supergiant's evolution with dredge-up of nuclear-processed
material are neglected. Certainly, all DD models assume that heating by
gravitational potential energy and He-burning provides expansion of the
envelope and creation (and maintenance) of a H-poor supergiant.
Clayton et al. (2007) suggested
that capture of the He white dwarf by the C-O white dwarf triggers an episode of
nucleosynthesis lasting several years. In this episode, the nuclear processes
include conversion of  $^{14}$N in the accreted material to $^{18}$O  but
destruction of  $^{18}$O by another $\alpha$-capture is considered  inhibited by
the cessation of conditions favorable for $\alpha$-capture. With a reaction
network activated by a `parametric' representation of the
temporal variation of the physical conditions,
 production of $^{18}$O was demonstrated as possible
during the merger. Clayton et al.'s (2007) abstract notes that `The merger
process is estimated to take only a few days, with accretion rates of 150
$M_\odot$ yr$^{-1}$ producing a temperature at the base of the accreted envelope
of 1.2$-$1.9 $\times 10^{8}$ K.' Nucleosynthesis at the envelope's base
continues for about 100 years. Invocation of $\alpha$-capture on $^{14}$N as the
source of $^{18}$O here, as in other models, requires that nucleosynthesis be
cut off before the $^{14}$N and $^{18}$O become exhausted.

\subsubsection{Model C}

The merger of the He white dwarf with the C-O white dwarf results in expansion
of the envelope to supergiant dimensions. This supergiant is initially hot
(i.e., an EHe or a hotter star) but quickly becomes a yellow supergiant (i.e.,
RCB or HdC). Evolution is terminated by a rapid crossing back to a hot
supergiant before descent of the white dwarf cooling track. Changes to the
surface composition may occur during life as a supergiant if the convective
envelope's base reaches sufficiently deep into the star.

Exploratory calculations of the effects of the dredge-up were undertaken by Saio
\& Jeffery (2002). In contrast to the rapid merger rate (150$M_\odot$ yr$^{-1}$)
adopted  by Clayton et al. (2007), Saio \& Jeffery adopted the much slower  rate
of 10$^{-5}M_\odot$ yr$^{-1}$: `approximately one-half of the Eddington limit
accretion rate for white dwarfs' (Saio \& Jeffery 2002). A Clayton et al. style
merger is complete in a few days but the post-merger nucleosynthesis is allowed
to continue for about 100 years.  In contrast,  accretion of (say) a
0.3$M_\odot$ He white dwarf takes 30,000 years at the Saio \& Jeffery rate.
Since accretion at this rate  suffices to drive the
transition from a white dwarf to a H-poor supergiant in much less than
30,000 years, the principal phase of
accretion as the H-poor  supergiant with the embedded merging pair of
stars is gaining in mass.

Illustrative calculations shown by Saio \& Jeffery (2002,  their Figures 3, 4,
and 5) for accretion of H-free material suggest that the supergiant's convective
envelope may not closely approach the He-burning shell in the initial
evolution from EHe to HdC. Just prior to the
final rapid evolution back to higher effective temperatures, a close
approach of the base of the convective envelope to the top of the
He-burning shell may introduce a difference in composition along the sequence
HdC-RCB-EHe. On the basis of these calculations, it would appear that the
surface abundances may not differ greatly from those provided by the cold merger
(models of type A), i.e., the convective envelope does not dredge up appreciable
amounts of material processed within the supergiant.    If H is included in the
accreted material, the convective envelope may encompass the H-burning shell in
the early phases of life as a cool H-poor supergiant with ensuing changes in
surface composition. Abundance of $^{18}$O  was not reported on by
Saio \& Jeffery (2002).

Obviously, additional calculations are needed to determine if and under what
conditions $^{18}$O may be produced at levels up to $^{18}$O/N$\sim0.3$. It will
be apparent that the ingredients
in such calculations are numerous (mass accretion
rate, mass and composition of the He white dwarf, mass and composition of the
outer layers of the C-O white dwarf that are mixed with the accreted material,
....). These ingredients define and influence the evolution of the
resulting H-poor supergiant and its surface composition including
changes resulting from dredge-up by its deep convective
envelope.  Much remains to be done!

\section{Concluding remarks}

On the observational front, the apparent difference in the $^{16}$O/$^{18}$O
ratios of the HdC and RCB stars is intriguing. Is this simply an artefact from
examining the necessarily small (complete) sample of HdC stars and a small, if
incomplete, sample of cool RCB stars? Or is the difference a clue to these
stars' origins? In order to shed light on these questions, it will be necessary
to find more of HdC stars and to observe more of the known cool RCBs. The total
sample of HdC stars has remained at five since the epochal paper by Warner
(1967). On the other hand, valuable additional data can be provided for RCBs.
High-resolution spectra of the first-overtone and especially the
fundamental CO bands should provide a more secure  estimate
of  the $^{16}$O/$^{18}$O ratio than those obtained by Clayton et al. (2007)
from medium-resolution spectra. Our result for S Aps
from first-overtone CO bands supports this contention.

Published  exploratory calculations  of the nucleosynthesis achieved during the
DD merger and subsequently deserve to be developed. One hopes that valuable
predictions will emerge about the range of and correlations between the C, N,
and O elemental and isotopic abundances and the abundance of other elements
affected by the nucleosynthesis, such as Li, F, and the trans-iron elements
produced by the $s$-process. As part of a test of predictions, it would be
desirable  to address the (different) carbon problems of the RCBs and HdCs.

Synthesis of fluorine occurs as indicated by observations of high F abundances
for EHe (Pandey 2006) and RCBs (Pandey, Lambert, \& Rao 2008). Clayton et al.
(2007) discuss F synthesis in their parametric modeling of a hot merger but the
maximum F abundance achieved is below the observed values. A particularly
fascinating challenge is offered by the presence of lithium in one HdC star (HD
148839, Rao \& Lambert 1996) and  four RCB stars (Asplund et al. 2000).
Conditions under which $^7$Li is produced  by the Cameron \& Fowler (1971)
mechanism from $^3$He are incompatible with those for synthesis of $^{18}$O from
$^{14}$N and F synthesis. Survival of Li in regions capable of synthesis of
$^{18}$O from $^{14}$N is impossible given that Li is destroyed by
$\alpha$-capture orders of magnitude faster than $^{14}$N is converted to
$^{18}$O via the intermediary $^{18}$F. Unfortunately, the Li-rich RCB stars are
among those for which the  first-overtone CO bands are  or are expected to be
absent or too weak for a determination of the O isotopic ratio. Our analysis of
HD 148839 offers the tantalising result that this Li-rich star may indeed have a
lower $^{18}$O abundance than other (Li-poor) HdC stars. High-resolution
infrared spectra of the CO fundamental of much higher S/N ratio would be of
interest.

The FF scenario may result in Li synthesis but is not expected to  account for
$^{18}$O-rich stars. Production of $^7$Li  is predicted (Herwig \& Langer 2001)
by the Cameron \& Fowler (1971) mechanism in which $^3$He is converted by
$\alpha$-capture and subsequent electron-capture to $^7$Li. In the DD scenario
which can potentially account for the $^{18}$O-rich stars, neither the He nor
the C-O  white dwarf can be expected to contain $^3$He, the necessary ingredient
for Li synthesis. Clayton et al. (2007) sketched how the nucleosynthesis may be
modified  by adding a H-rich envelope (presumably a potential reservoir of
$^3$He) to the He white dwarf. Their sketch did not cover lithium synthesis. 
As in other areas of stellar astrophysics, lithium is at the center of very
fascinating questions of stellar nucleosynthesis and evolution.

\acknowledgments{This paper is based on observations obtained at the Gemini
Observatory, which is operated by the Association of Universities for Research
in Astronomy, Inc., under a cooperative agreement with the NSF on behalf of the
Gemini partnership: the National Science Foundation (United States), the Science
and Technology Facilities Council (United Kingdom), the National Research
Council (Canada), CONICYT (Chile), the Australian Research Council (Australia),
Minist\'{e}rio da Ci\^{e}ncia e Tecnologia (Brazil) and SECYT (Argentina). The
observations were obtained with the Phoenix infrared spectrograph, which was
developed and is operated by the National Optical Astronomy Observatory.  The
spectra were obtained as part of programs GS-2006A-C-13 and GS-2007A-DD-1. This
research has been supported in part by the Robert A. Welch Foundation of
Houston, Texas. KE gratefully acknowledges support from the Swedish Research
Council.}

\clearpage

\begin{deluxetable}{lcccc}
\tabletypesize{\scriptsize}
\tablecaption{Observation summary of the near-IR PHOENIX observations$^{a}$\label{tbl-1}}
\tablewidth{0pt}
\tablehead{
\colhead{Star} & \colhead{Spectral ranges observed}  & \colhead{Date}  & \colhead{Spectral ranges observed} & \colhead{Date} \\
\colhead{} & \colhead{$\mu$m}  & \colhead{}  & \colhead{$\mu$m} & \colhead{} \\
}
\startdata
HD 137613         & 2.332, 2.343, 2.354, 2.366        & 06/18/06 & 2.251, 2.354 & 02/01/07 \\
HD 175893         & 2.251, 2.332, 2.343, 2.354, 2.366 & 06/18/06 &$\dots$ &  $\dots$  \\
HD 182040         & 2.251, 2.332, 2.343, 2.354, 2.366 & 06/18/06 &$\dots$ &  $\dots$  \\
HD 173409         & 2.332, 2.343, 2.354, 2.366        & 06/18/06 &$\dots$ &  $\dots$  \\
HD 148839         & 2.332, 2.343, 2.354, 2.366        & 06/18/06 & 2.251, 2.354 & 02/01/07 \\
S Aps             &$\dots$                     & $\dots$  & 2.251, 2.354 & 02/01/07 \\
UW Cen            &$\dots$                     & $\dots$  & 2.251, 2.354 & 02/01/07, 02/07/07 \\
Y Mus             &$\dots$                     & $\dots$  & 2.251, 2.354 & 02/01/07 \\
V854 Cen          &$\dots$                     & $\dots$  & 2.251, 2.354 & 02/01/07, 02/07/07 \\
\enddata

\tablenotetext{a}{The first five objects (with HD names) are the five HdC stars known
while the rest of stars are the four RCB stars observed.}
\end{deluxetable}

\begin{deluxetable}{lccccccccc}
\tabletypesize{\scriptsize}
\tablecaption{Derived chemical abundances in HdC and RCB stars$^{a}$\label{tbl-2}}
\tablewidth{0pt}
\tablehead{
\colhead{Star} &\colhead{T$_{\rm eff}$ (K)$^{b}$} &\colhead{$\xi$ (km s$^{-1}$)} &
\colhead{$^{16}$O/$^{18}$O}  & \colhead{$^{16}$O/$^{17}$O} &
\colhead{$^{14}$N/$^{15}$N} & \colhead{$^{12}$C/$^{13}$C} & \colhead{C/N/O$^{c}$} &
\colhead{$<$Na$>$} & \colhead{$<$S$>$} \\
}
\startdata
HD 137613 &5400 & 7 & 0.5   & $>$50 & $>$10   & $>$10 & 9.5/9.4/8.7       & 6.9  & 7.6  \\
HD 175893 &5500 & 7 & 0.3   & $>$50 & $>$10   & $>$10 & 9.5/9.2/8.7       & 6.5  & 7.6  \\
HD 182040 &5600&  7&0.5   & $>$8 & $>$10& $>$10 & 9.5/9.2/8.0       &6.8  & 7.4  \\
HD 173409 &6100& 7 &$\dots$ & $\dots$& $\dots$ & $\dots$&9.5/9.2/8.7& 7.0     & $\dots$    \\
HD 148839 &6500 & 7 & $\dots$  & $\dots$  & $\dots$ &  $\dots$& 9.5/9.8/9.2  & 7.0   & 7.4    \\
S Aps     &5400& 7 &16    & $>$80 & $>$10   &$\dots$& 9.5/9.6/9.4       &$\dots$& 6.8    \\
Y Mus     &7200& 10&$\dots$&$\dots$& $\dots$    &$\dots$& 9.5/9.4/$\dots$  &$\dots$& 6.8    \\     
UW Cen$^{d}$    &7400& 12&$\dots$&$\dots$& $\dots$  &$\dots$& 9.5/$<$9.8/$\dots$&$\dots$& 6.0    \\     
V854 Cen$^{d}$  &6750   & 6 &$\dots$     &$\dots$       &$\dots$       &$\dots$      &$\dots$  &  $\dots$   & $\dots$ \\     
\enddata

\tablenotetext{a}{The abundances are normalized to log $\sum \mu_i
\epsilon_{i}$=12.15, i.e. log$\varepsilon$(He)=11.52 and log$\varepsilon$(C)=9.52.}
\tablenotetext{b}{References.- Clayton et al. (2007); Asplund et al. (1997, 1998); Lawson et al. (1990).}
\tablenotetext{c}{The CNO abundances are given for the assumed input composition
i.e. log$\varepsilon$(Fe)=7.2, log$\varepsilon$(Na)=6.8, log$\varepsilon$(S)=7.5, etc. Asplund et al. 1997)}
\tablenotetext{d}{The K-band spectrum is largely from circumstellar dust and the
photospheric spectrum is greatly obscured or diluted.}
\end{deluxetable}

\begin{deluxetable}{ccccccc}
\tabletypesize{\scriptsize}
\tablecaption{Sensitivity of the derived $^{16}$O/$^{18}$O ratio and N, O, Na and S abundances to slight
changes in the model atmosphere parameters for the HdC star HD 175893\label{tbl-3}}
\tablewidth{0pt}
\tablehead{
\colhead{Adopted value} &\colhead{Change}& \colhead{$\Delta$$^{16}$O/$^{18}$O} & \colhead{$\Delta$log$\epsilon$(N)} & \colhead{$\Delta$log$\epsilon$(O)}& \colhead{$\Delta$log$\epsilon$(Na)} & \colhead{$\Delta$log$\epsilon$(S)}
}
\startdata
\textit{T$_{\rm eff}$}=5500 K & $\Delta$\textit{T$_{\rm eff}$}=$\pm$100 K & $\mp$0.1 & $\pm$0.1 & $\pm$0.1& $\pm$0.05& $\pm$0.05\\
\textit{log g}=$+$0.5 & $\Delta$\textit{log g}=$\pm$1.0 & $\mp$0.1  & $-$0.2/$+$0.4 & $-$0.1/$+$0.2&$-$0.2/$+$0.7 & $-$0.2/$+$0.4\\
\textit{$\xi$}=7 km s$^{-1}$&$\Delta$\textit{$\xi$}=$\pm$2 km s$^{-1}$&$\pm$0.2 & $\pm$0.05 & $\pm$0.05 & $\pm$0.1& $\pm$0.1\\
\textit{FWHM}=6 km s$^{-1}$&$\Delta$\textit{FWHM}=$\pm$2 km s$^{-1}$&$\mp$0.05& $\pm$0.05& $\pm$0.05 & $\pm$0.1& $\pm$0.1\\
\enddata
\end{deluxetable}

\clearpage

\begin{figure}
\includegraphics[angle=-90,scale=.60]{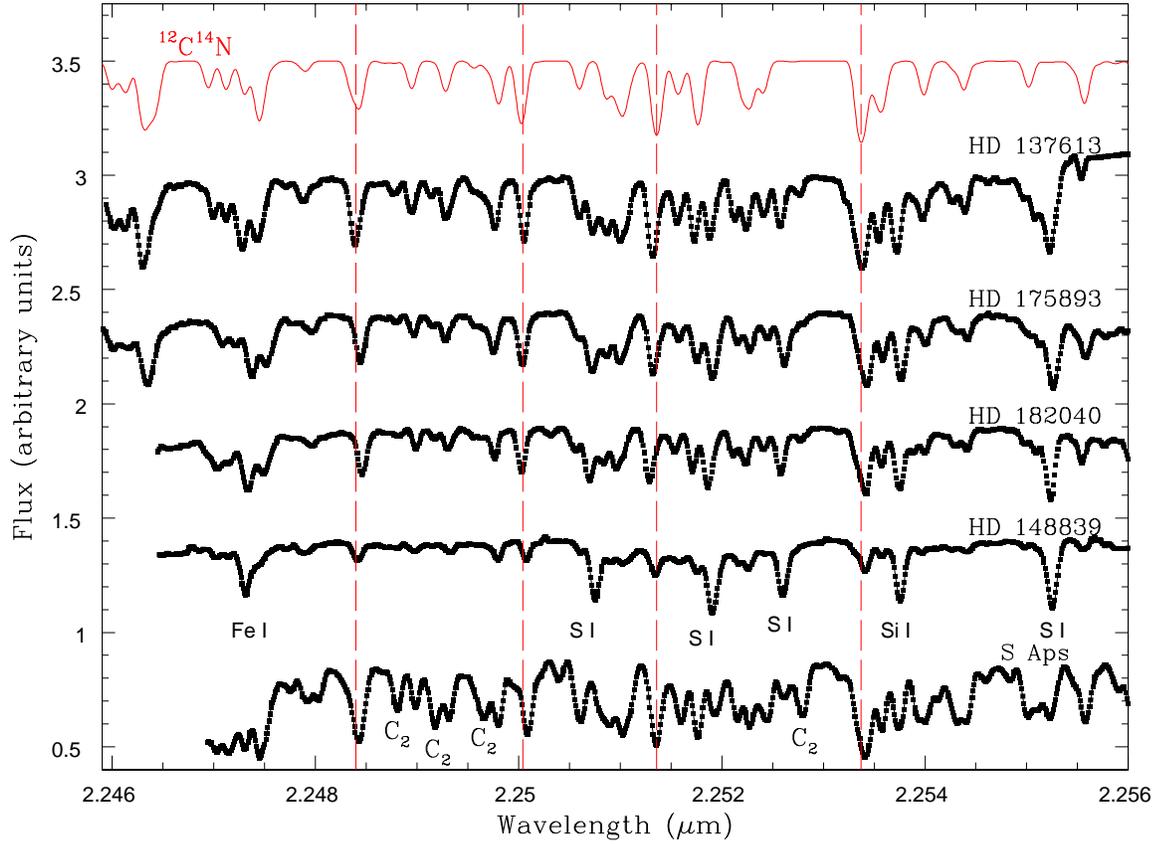}
\caption{PHOENIX spectra centered at $\sim$2.251 $\mu$m for four HdC stars (from
top to bottom: HD 137613, HD 175893, HD 182040, and HD 148839, respectively).
The spectrum of the RCB star S Aps is also shown. Positions  of several  atomic
lines  are labeled. The strongest $^{12}$C$^{14}$N features are indicated with a
dashed red vertical line. A $^{12}$C$^{14}$N synthetic spectrum (in red)
composed for HD 137613 is  shown at the top. Phillips system C$_2$ lines labeled
are 1-3 R22, 0-2 Q34, 1-3 Q8 and 1-3 Q10 in order of increasing wavelength.
\label{fig1}}
\end{figure}

\clearpage

\begin{figure}
\includegraphics[angle=-90,scale=.60]{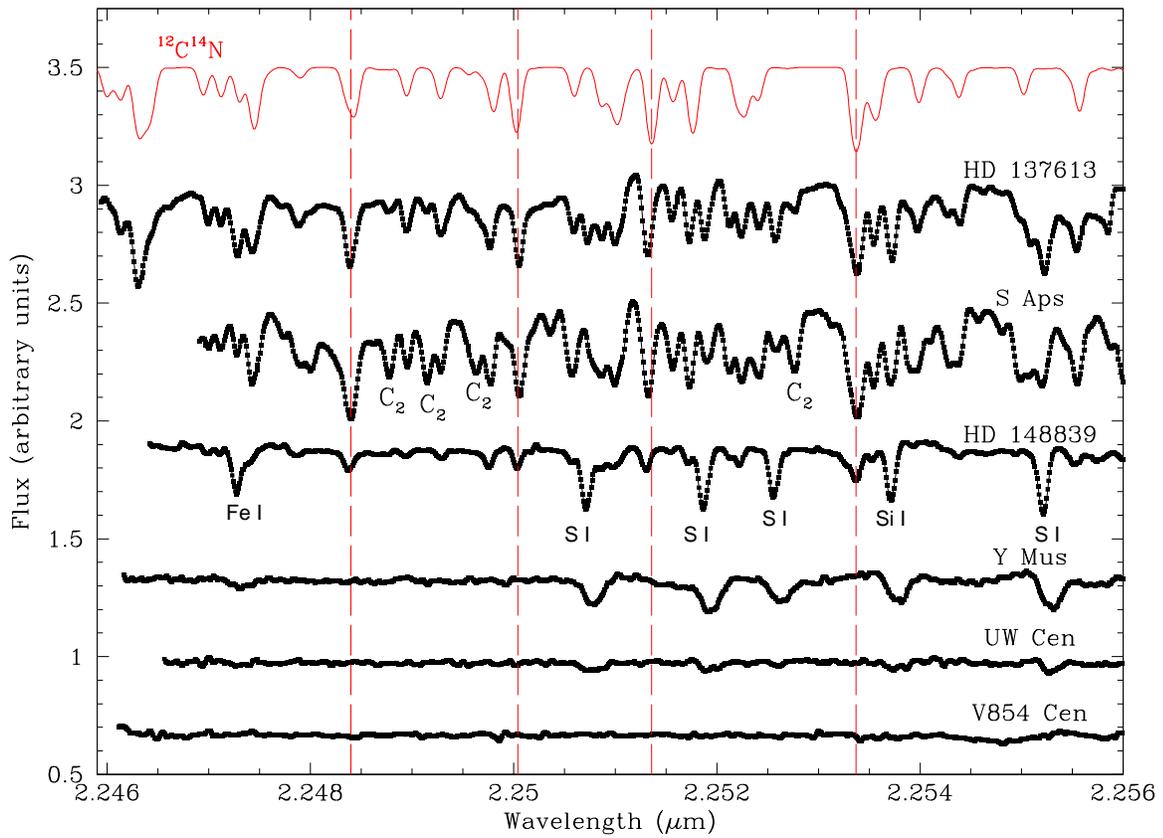}
\caption{PHOENIX spectra centered at $\sim$2.251 $\mu$m
 for  four RCB stars and two HdC stars (from top to
bottom: HD 137613, S Aps, HD 148839, Y Mus, UW Cen, and V854 Cen,
respectively). See caption of Figure 1 for
additional information. \label{fig2}}
\end{figure}

\clearpage

\begin{figure}
\includegraphics[angle=-90,scale=.60]{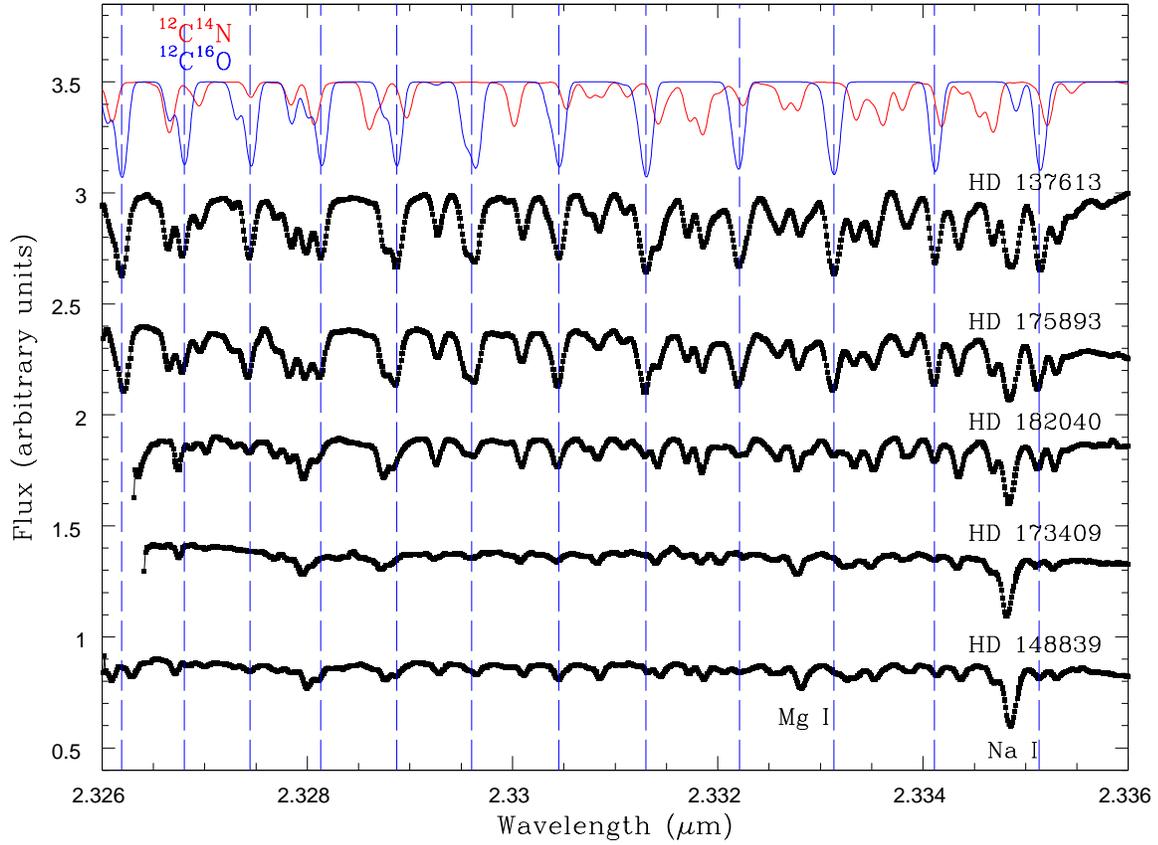}
\caption{PHOENIX spectra centered at 2.331 $\mu$m
for the five HdC stars. This
region is dominated by  lines of the $^{12}$C$^{16}$O and $^{12}$C$^{14}$N
molecules but note the Na\,{\sc i} line at 2.3348 $\mu$m. The strongest
$^{12}$C$^{16}$O  lines are marked with a dashed blue vertical line. Synthetic
spectra composed for HD 137613 for the CO and CN isomers ($^{12}$C$^{16}$O in
blue and $^{12}$C$^{14}$N in red) shown at the top indicate that these molecules
dominate the spectra. \label{fig3}}
\end{figure}

\clearpage

\begin{figure}
\includegraphics[angle=-90,scale=.60]{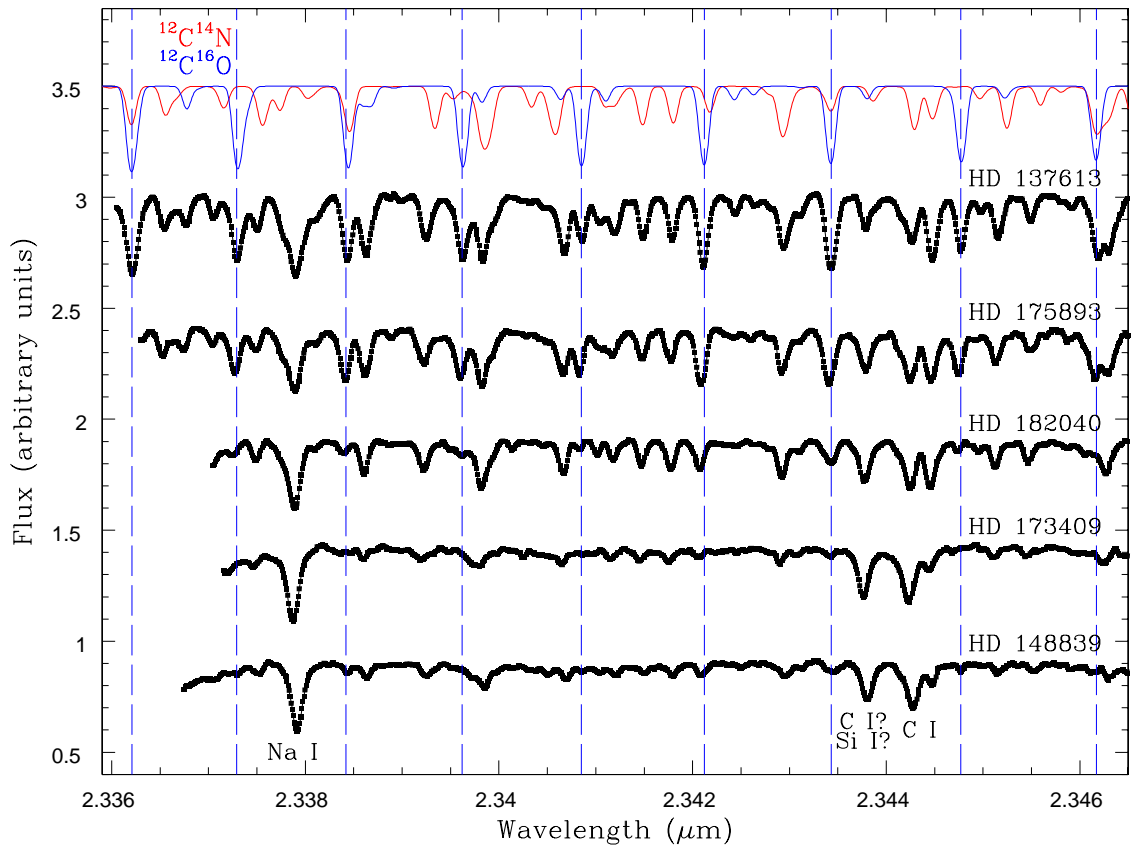}
\caption{PHOENIX spectra centered at  2.341 $\mu$m
 for the five  HdC stars. As
for  Figure 3, this region is  dominated by $^{12}$C$^{16}$O and
$^{12}$C$^{14}$N lines but note the Na\,{\sc i} line at 2.3379 $\mu$m. See the
caption of Figure 3 for additional information. \label{fig4}}
\end{figure}

\clearpage

\begin{figure}
\includegraphics[angle=-90,scale=.60]{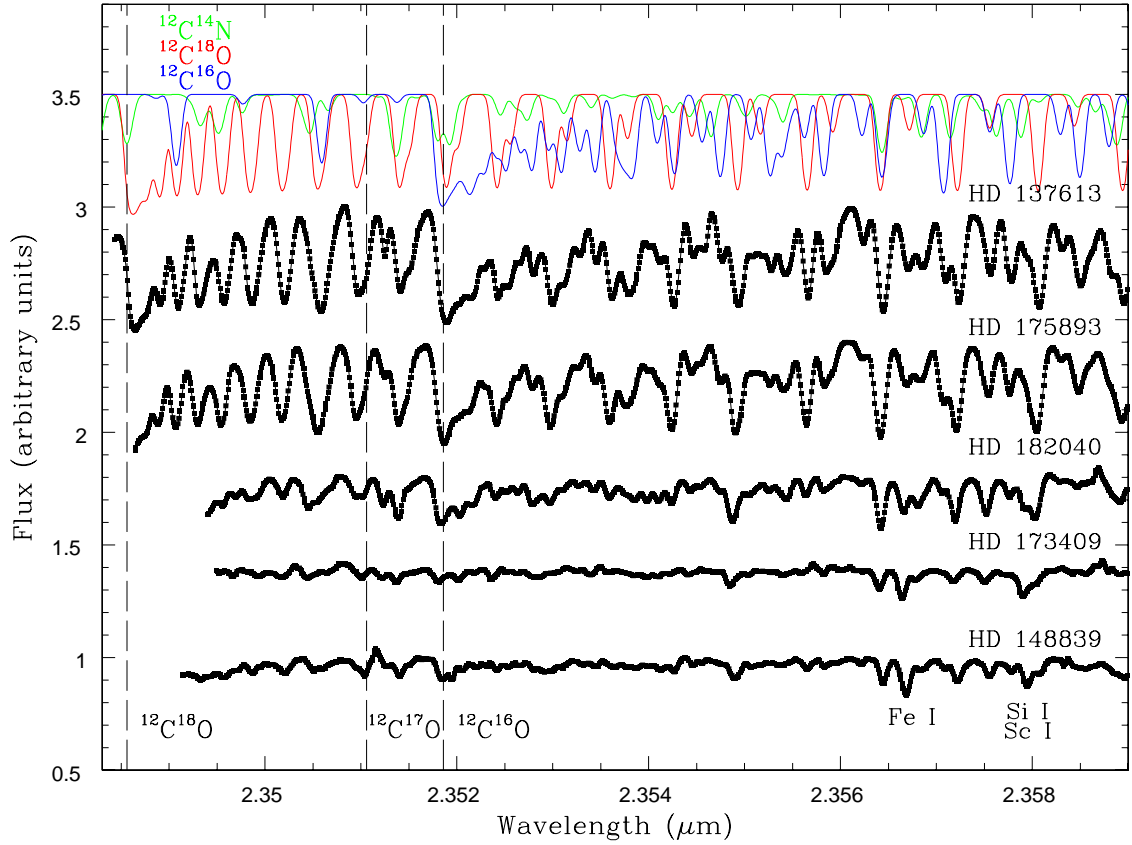}
\caption{PHOENIX spectra  centered at $\sim$2.354 $\mu$m
for the five known HdC stars (from top to bottom: HD
137613, HD 175893, HD 182040, HD 173409, and HD 148839, respectively).
 Wavelengths  of 2-0 $^{12}$C$^{18}$O, 3-1 $^{12}$C$^{17}$O,
and the 4-2 $^{12}$C$^{16}$O bandheads are marked with a vertical dashed line.
Synthetic spectra composed for HD 137613 are shown for the isomers $^{12}$C$^{16}$O
in blue, $^{12}$C$^{18}$O in red and $^{12}$C$^{14}$N in green. \label{fig5}}
\end{figure}

\clearpage

\begin{figure}
\includegraphics[angle=-90,scale=.60]{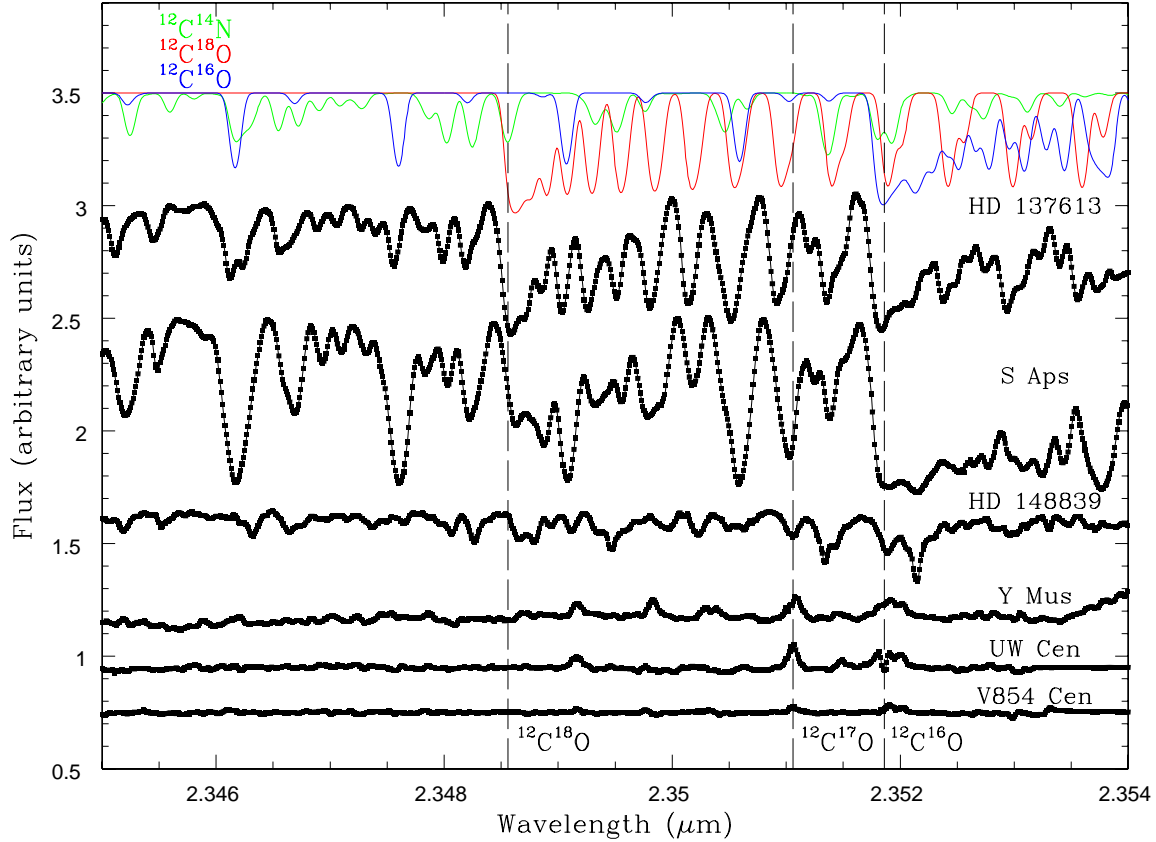}
\caption{PHOENIX spectra around the 2-0 $^{12}$C$^{18}$O at $\sim$2.3485 $\mu$m
 for  six stars (four RCB stars and two HdC stars): (from
top to bottom) HD 137613, S Aps, HD 148839, Y Mus, UW Cen, and V854 Cen.
 Synthetic spectra composed for HD
137613 for the CO isomers ($^{12}$C$^{16}$O in blue and $^{12}$C$^{18}$O in red)
and for the $^{12}$C$^{14}$N isomer (in green) are shown at the top for
comparison. The apparent emission features in the spectra of the Y Mus, UW Cen,
and V854 Cen arise from incomplete cancelling of telluric lines.  See the caption
of Figure 5 for additional information. \label{fig6}}
\end{figure}

\clearpage

\begin{figure}
\includegraphics[angle=-90,scale=.60]{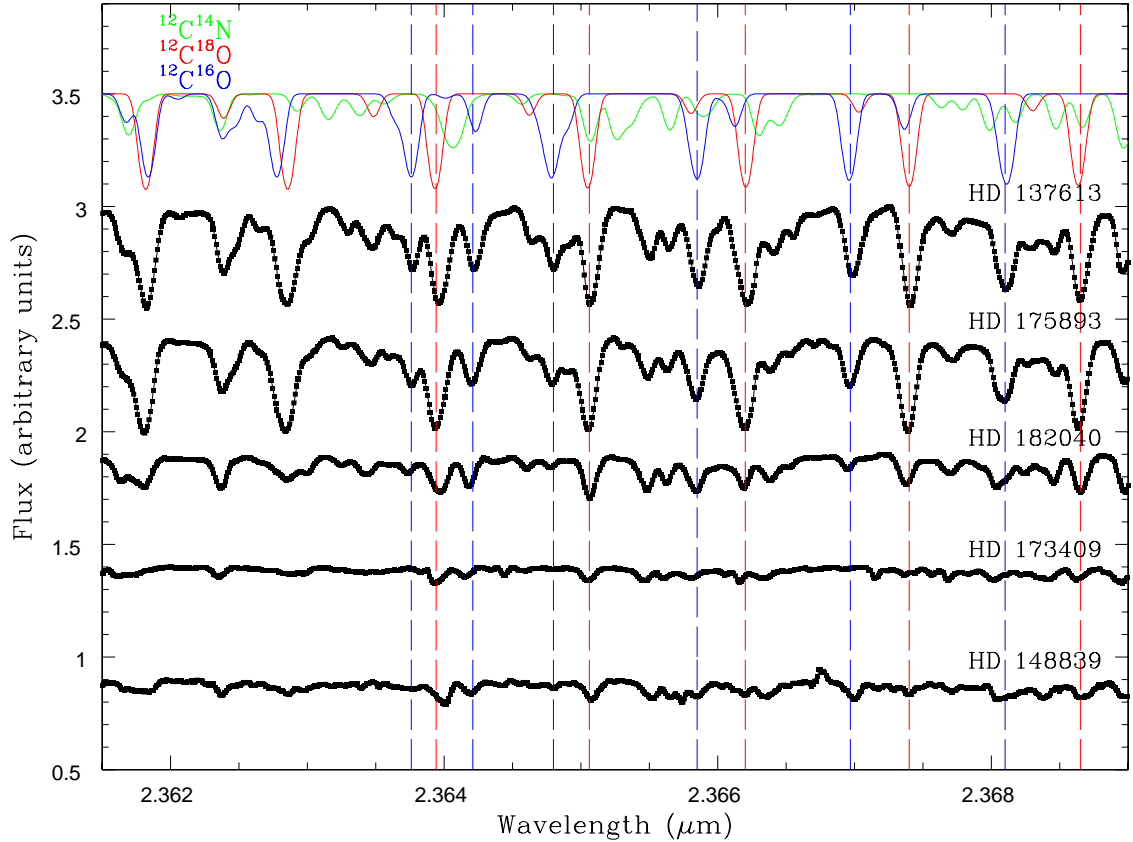}
\caption{PHOENIX spectra centered at 2.365 $\mu$m
 of  the five known HdC stars.
Synthetic spectra composed for HD 137613 for the CO isomers ($^{12}$C$^{16}$O in
blue and $^{12}$C$^{18}$O in red) and for the $^{12}$C$^{14}$N isomer (in green)
are shown at the top for comparison. Note that individual isotopic molecular lines
of $^{12}$C$^{18}$O and $^{12}$C$^{16}$O are clearly resolved and marked with
dashed vertical lines. \label{fig7}}
\end{figure}

\clearpage

\begin{figure}
\includegraphics[angle=-90,scale=.60]{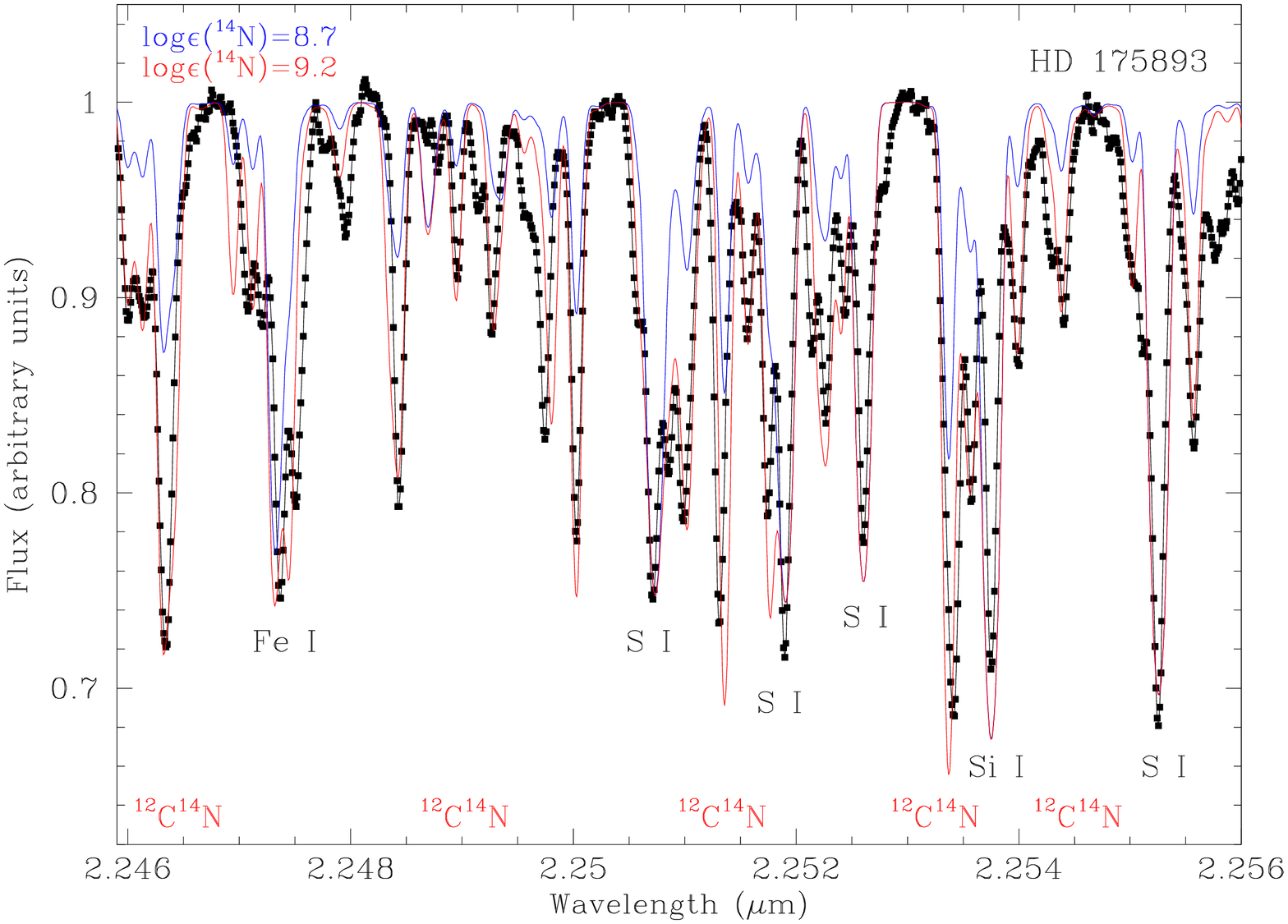}
\caption{Best synthetic  (red) and observed  (black) spectrum for the CN region
centered at $\sim$2.251 $\mu$m for the $^{18}$O-rich HdC star HD 175893. The
synthetic spectrum has been created
assuming the input C abundance log$\varepsilon$(C)=9.52,
the  N is in the form of $^{14}$N with an abundance log$\varepsilon$(N)=9.2,
and a total O abundance of log$\varepsilon$(O)= 8.7.  A synthetic spectrum
(blue) for a lower $^{14}$N abundance of log$\varepsilon$(N)=8.7 dex is  shown
for comparison. Note the strong dependence of the $^{12}$C$^{14}$N lines to
changes of the elemental N abundance.
 \label{fig8}}
\end{figure}

\clearpage

\begin{figure}
\includegraphics[angle=-90,scale=.60]{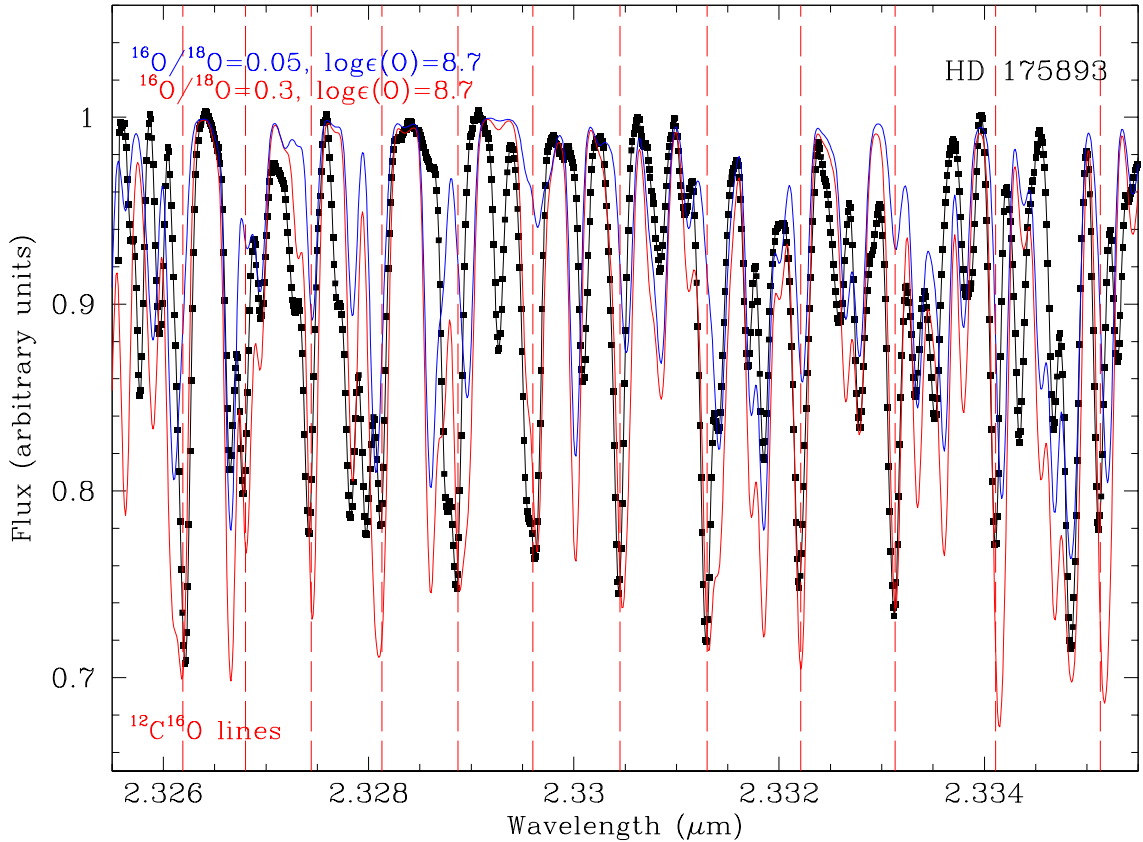}
\caption{Best synthetic (red) and observed (black) spectrum for the wavelength
region centered around 2.332 $\mu$m for the $^{18}$O-rich HdC star HD 175893.
The red spectrum assumes the input C abundance of 9.52, the N abundance of 9.2, a total
O ($^{16}$O $+$ $^{18}$O) of log$\varepsilon$(O)=8.7, and the isotopic ratio
$^{16}$O/$^{18}$O=0.3. This region contains $^{12}$C$^{16}$O but not
$^{12}$C$^{18}$O lines and the $^{12}$C$^{16}$O lines are fit with the isotopic
ratio $^{16}$O/$^{18}$O=0.3 (or log$\varepsilon$($^{16}$O)=8.1). The synthetic
spectrum also includes CN lines. A synthetic spectrum (blue) for a lower
$^{16}$O/$^{18}$O ratio of 0.05 (or a lower $^{16}$O abundance) but the same
total O ($^{16}$O $+$ $^{18}$O) of log$\varepsilon$(O)=8.7 is shown for
comparison. Note the strong dependence of the most clear $^{12}$C$^{16}$O lines
(marked with red dashed vertical lines) to changes of the $^{16}$O abundance.\label{fig9}}
\end{figure}

\clearpage

\begin{figure}
\includegraphics[angle=-90,scale=.60]{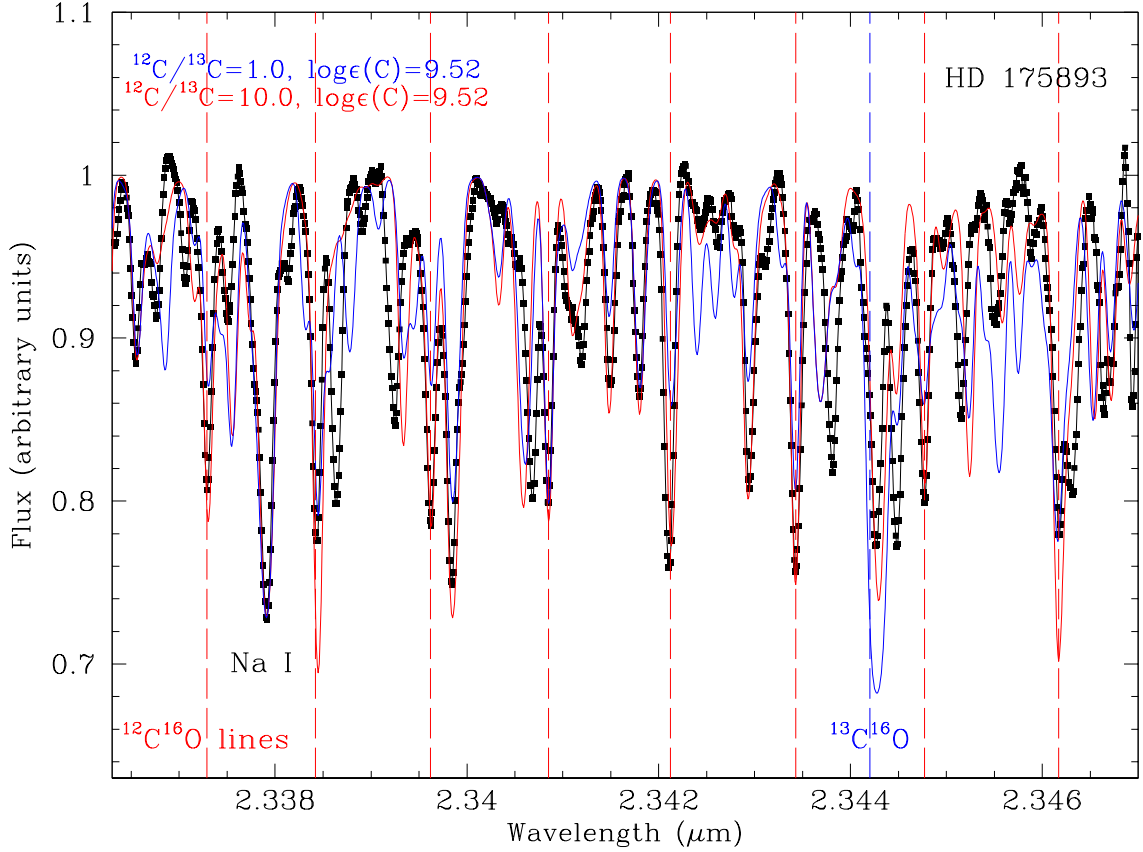}
\caption{Best synthetic (red) and observed (black) spectrum in the region around
the 2.344 $\mu$m $^{13}$C$^{16}$O bandhead for the $^{18}$O-rich HdC star HD
175893. The red spectrum assumes the input C abundance of 9.52, the N abundance
of 9.2, a total O ($^{16}$O $+$ $^{18}$O) of log$\varepsilon$(O)=8.7, and the
isotopic ratios $^{16}$O/$^{18}$O and $^{12}$C/$^{13}$C of 0.3 and 10.0,
respectively. This region contains $^{12}$C$^{16}$O but not $^{12}$C$^{18}$O
lines. The 2-0 $^{13}$C$^{16}$O bandhead is indicated with a blue dashed
vertical line. The synthetic spectrum also includes CN lines. A synthetic
spectrum (blue) for a lower $^{12}$C/$^{13}$C ratio of 1.0 (or a higher $^{13}$C
abundance) but the same total C ($^{12}$C $+$ $^{13}$C) of
log$\varepsilon$(C)=9.52 is shown for comparison. Note the strong dependence of
the 2-0 $^{13}$C$^{16}$O bandhead to changes of the $^{12}$C/$^{13}$C ratio.
\label{fig10}}
\end{figure}

\clearpage

\begin{figure}
\includegraphics[angle=-90,scale=.60]{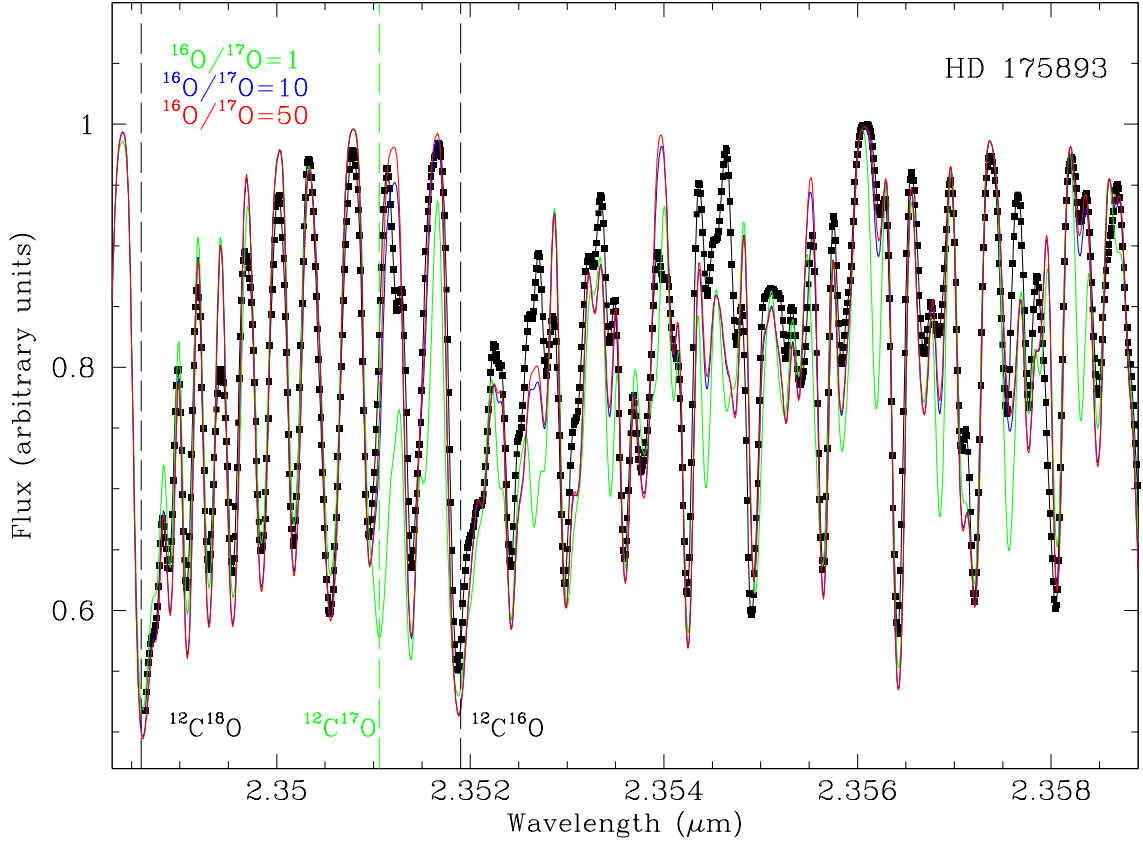}
\caption{Best synthetic (red) and observed (black) spectrum in the region around
the 2-0 2.349$\mu$m $^{12}$C$^{18}$O bandhead for the $^{18}$O-rich HdC star HD
175893.
 The spectrum also covers the expected
position of the 3-1 2.3511$\mu$m $^{12}$C$^{17}$O bandhead.
 The red spectrum assumes the input C abundance of 9.52, the N abundance of 9.2,
a total O ($^{16}$O $+$ $^{17}$O $+$ $^{18}$O) of log$\varepsilon$(O)=8.7 and
the isotopic ratios $^{16}$O/$^{18}$O=0.3 and $^{16}$O/$^{17}$O=50. This region
contains $^{12}$C$^{16}$O and $^{12}$C$^{18}$O lines which are fit with the
isotopic ratio $^{16}$O/$^{18}$O=0.3.
 Synthetic spectra
for lower $^{16}$O/$^{17}$O ratios (or higher $^{17}$O abundances) of 1 (green)
and 10 (blue) but the same total O ($^{16}$O $+$ $^{17}$O $+$ $^{18}$O) of
log$\varepsilon$(O)=8.7 are shown for comparison. Note the strong dependence of
the 3-1 2.3511$\mu$m $^{12}$C$^{17}$O bandhead (not detected in our spectra) to
changes of the $^{16}$O/$^{17}$O ratio. \label{fig11}}
\end{figure}

\clearpage

\begin{figure}
\includegraphics[angle=-90,scale=.60]{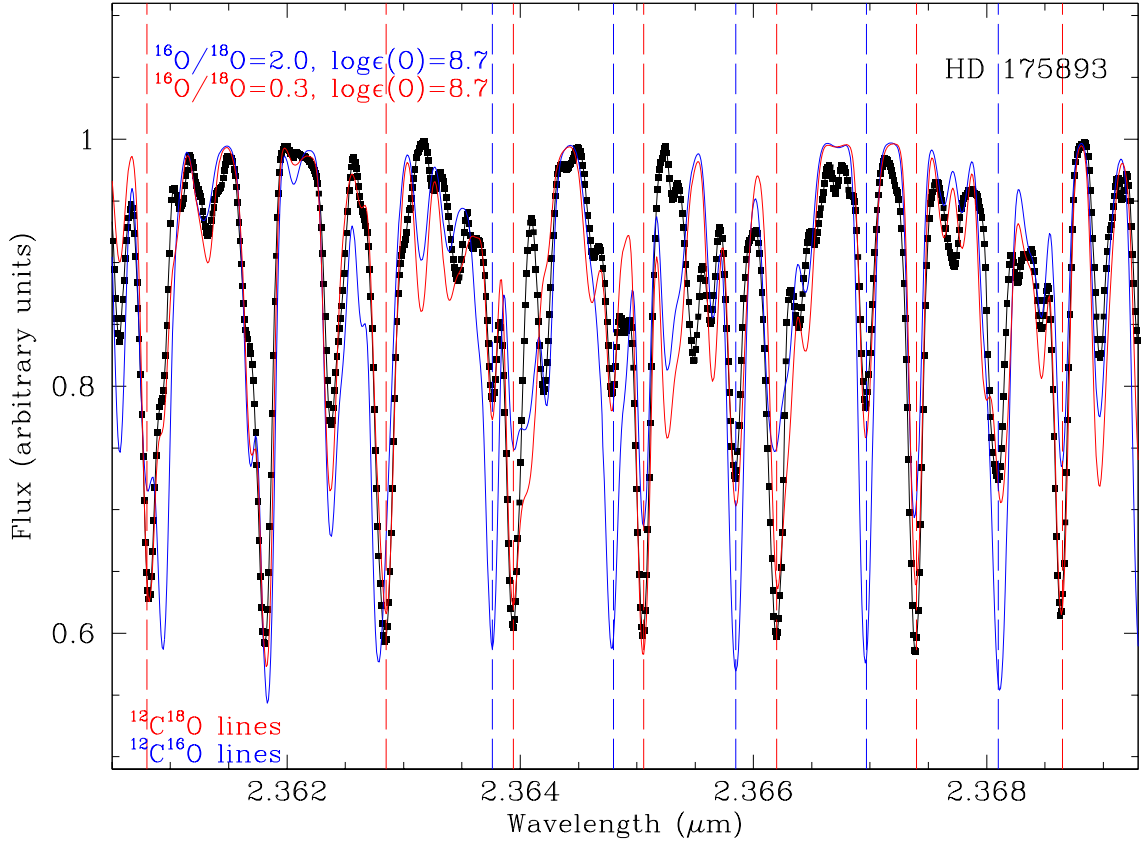}
\caption{Best synthetic (red)  and observed (black) spectrum for the region
centered at 2.366 $\mu$m for the $^{18}$O-rich HdC star HD 175893. The red
spectrum assumes the input C abundance of 9.52, the N abundance of 9.2,  a total O
($^{16}$O $+$ $^{18}$O) of log$\varepsilon$(O)=8.7 and the isotopic ratio
$^{16}$O/$^{18}$O=0.3. The most clear $^{12}$C$^{18}$O and $^{12}$C$^{16}$O
lines are marked with red and blue vertical dashed lines, respectively. A
synthetic spectrum (blue) for a higher $^{16}$O/$^{18}$O ratio of 2 (or a lower
$^{18}$O abundance) but the same  total O ($^{16}$O $+$ $^{18}$O) of
log$\varepsilon$(O)=8.7 is shown for comparison. \label{fig12}}
\end{figure}

\clearpage

\begin{figure}
\includegraphics[angle=-90,scale=.60]{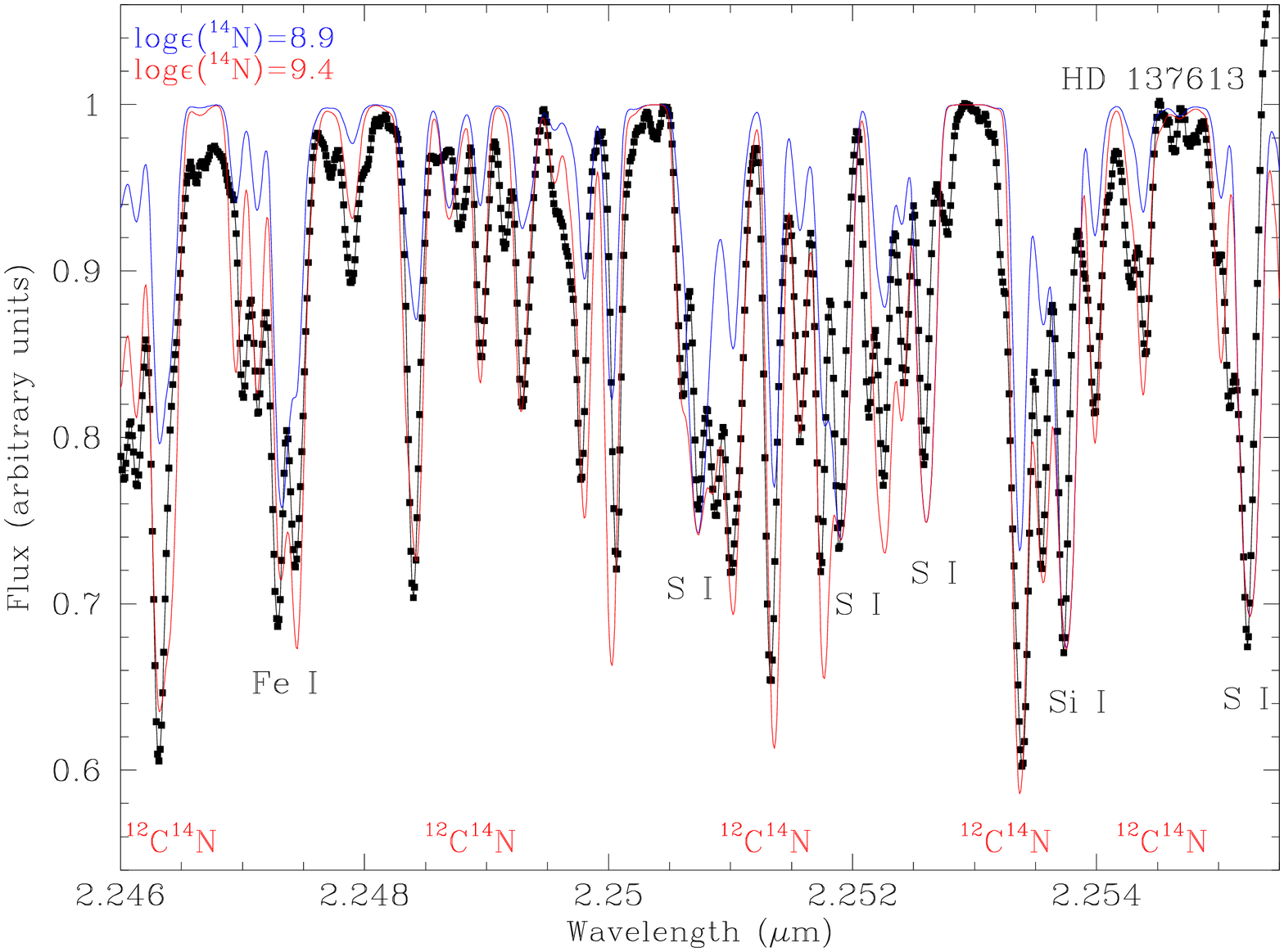}
\caption{Best synthetic (red) and observed (black) spectrum for the CN region
centered at $\sim$2.251 $\mu$m for the $^{18}$O-rich HdC star HD 137613. The
synthetic spectrum has been created  assuming the input C abundance of 9.52,
the N is in the form of $^{14}$N with an abundance log$\varepsilon$(N)=9.4 dex,
and a total O abundance of log$\varepsilon$(O)= 8.7.  A synthetic spectrum
(blue) for a lower $^{14}$N abundance of log$\varepsilon$(N)=8.9 dex is also shown
for comparison. \label{fig13}}
\end{figure}

\clearpage

\begin{figure}
\includegraphics[angle=-90,scale=.60]{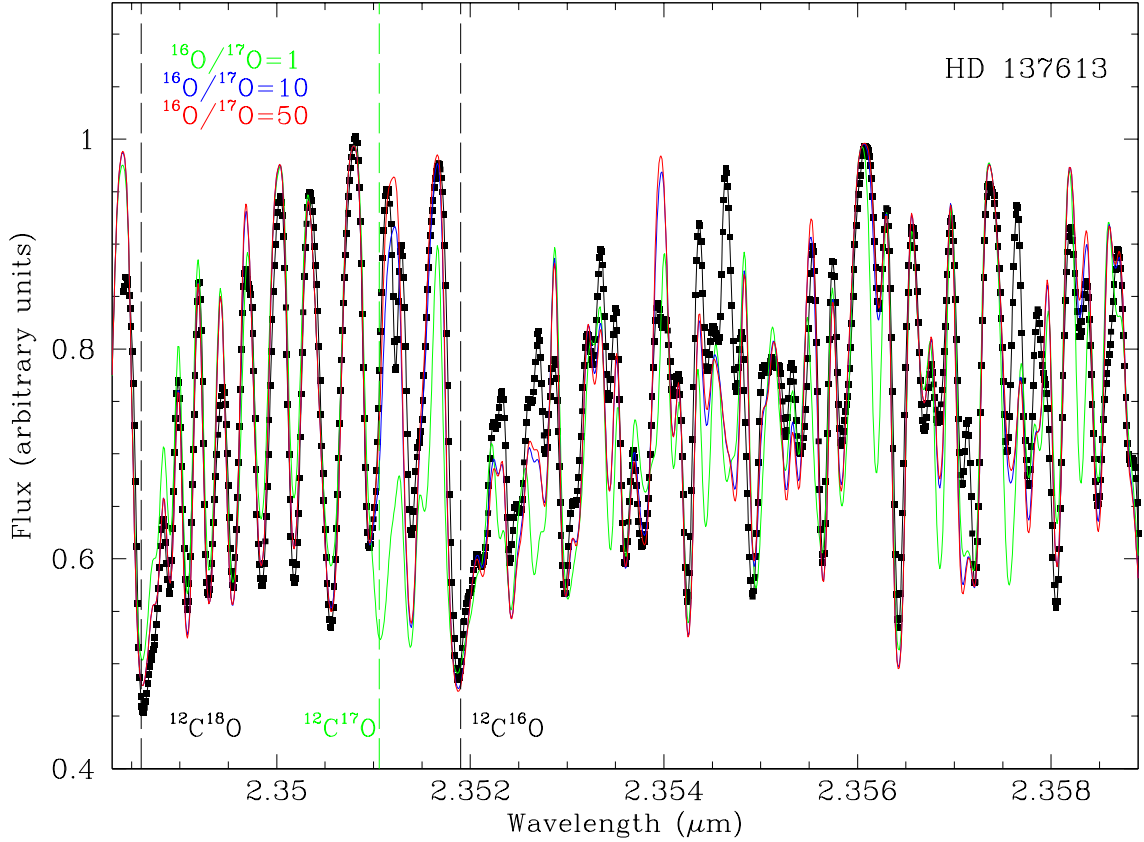}
\caption{Best synthetic (red) and observed (black) spectrum in the region around
the 2-0 2.349$\mu$m $^{12}$C$^{18}$O bandhead for the $^{18}$O-rich HdC star HD
137613. The red spectrum assumes the input C abundance of 9.52, the N abundance of 9.4,
a total O ($^{16}$O $+$ $^{17}$O $+$ $^{18}$O) of log$\varepsilon$(O)=8.7 and
the isotopic ratios $^{16}$O/$^{18}$O=0.5 and $^{16}$O/$^{17}$O=50. Synthetic
spectra for lower $^{16}$O/$^{17}$O ratios (or higher $^{17}$O abundances) of 1
(green) and 10 (blue) but the same total O ($^{16}$O $+$ $^{17}$O $+$ $^{18}$O)
of log$\varepsilon$(O)=8.7 are shown for comparison. See Figure 11 for
additional information. \label{fig14}}
\end{figure}

\clearpage

\begin{figure}
\includegraphics[angle=-90,scale=.60]{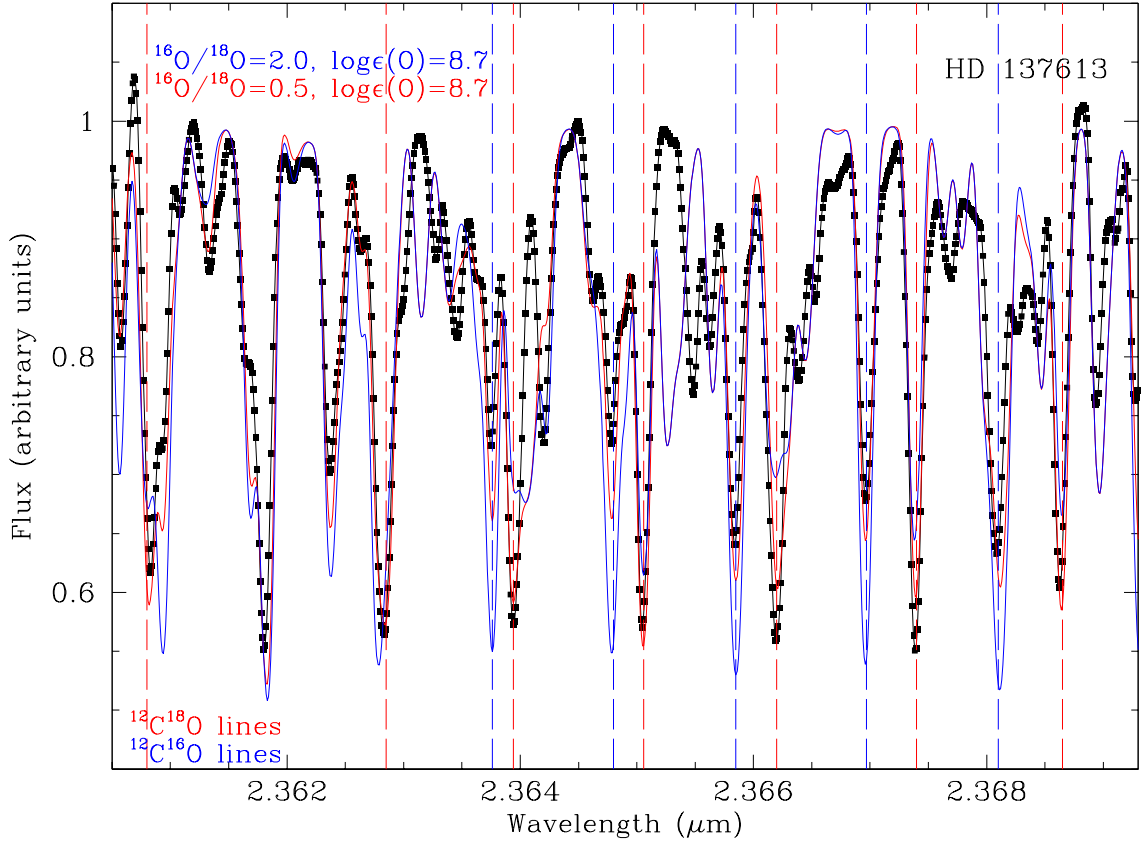}
\caption{Best synthetic (red)  and observed (black) spectrum for the region
centered at 2.366 $\mu$m for the $^{18}$O-rich HdC star HD 137613. The red
spectrum assumes the input C abundance of 9.52, the N abundance of 9.4, a total O
($^{16}$O $+$ $^{18}$O) of log$\varepsilon$(O)=8.7 and the isotopic ratio
$^{16}$O/$^{18}$O=0.5. Some of the strongest $^{12}$C$^{18}$O and
$^{12}$C$^{16}$O lines are marked with red and blue vertical dashed lines,
respectively. A synthetic spectrum (blue) for a higher $^{16}$O/$^{18}$O ratio
of 2 (or a lower $^{18}$O abundance) but the same  total O ($^{16}$O $+$
$^{18}$O) of log$\varepsilon$(O)=8.7 is shown for comparison.\label{fig15}}
\end{figure}

\clearpage

\begin{figure}
\includegraphics[angle=-90,scale=.60]{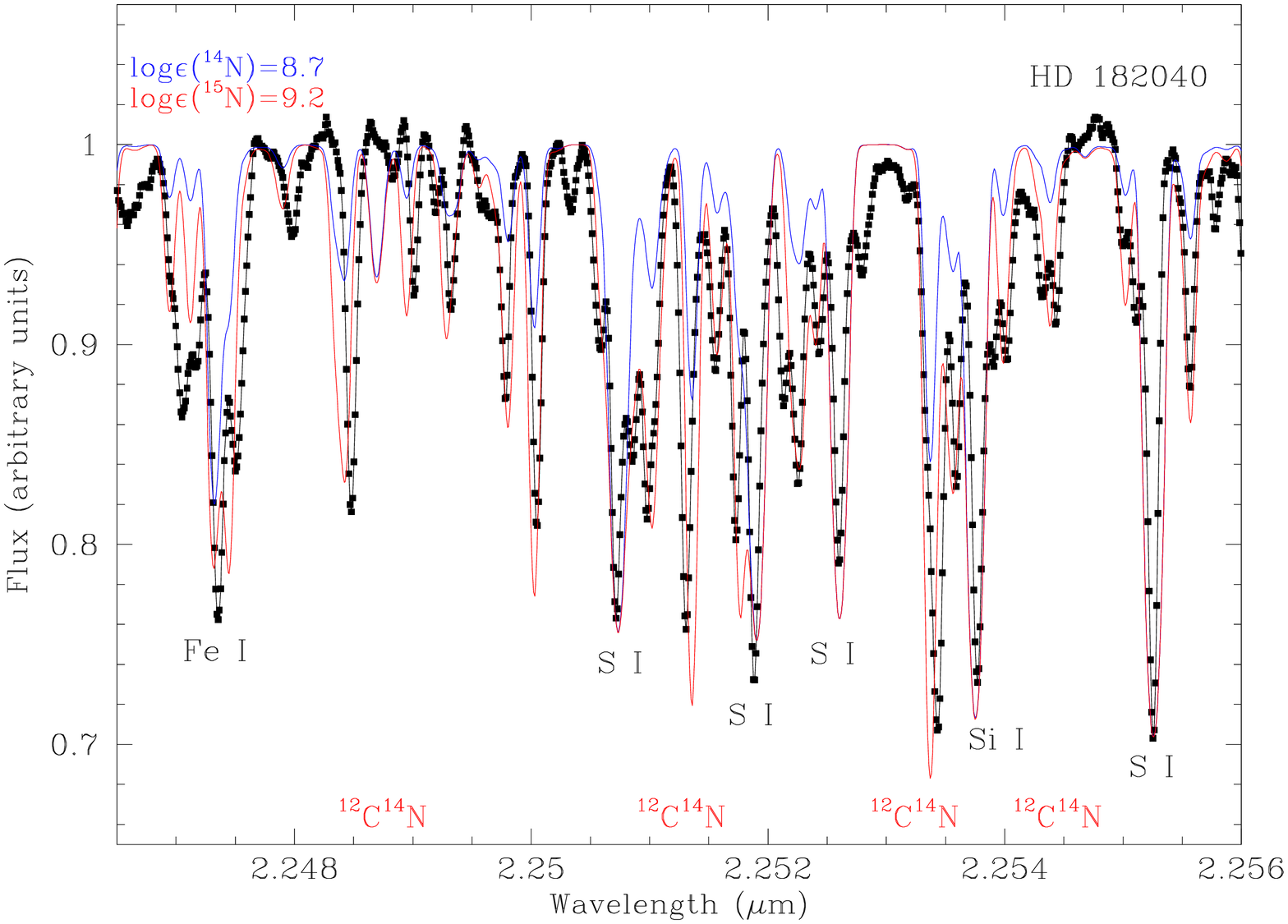}
\caption{Best synthetic (red) and observed (black) spectrum for the CN region
centered at $\sim$2.251 $\mu$m for the $^{18}$O-rich HdC star HD 182040. The
synthetic spectrum has been created assuming the input C abundance of 9.52,
the N is in the form of $^{14}$N with an abundance log$\varepsilon$(N)=9.2 dex,
and a total O abundance of log$\varepsilon$(O)= 8.0. A synthetic spectrum
(blue) for a lower $^{14}$N abundance of log$\varepsilon$(N)=8.7 dex is also
shown for comparison. \label{fig16}}
\end{figure}

\clearpage

\begin{figure}
\includegraphics[angle=-90,scale=.60]{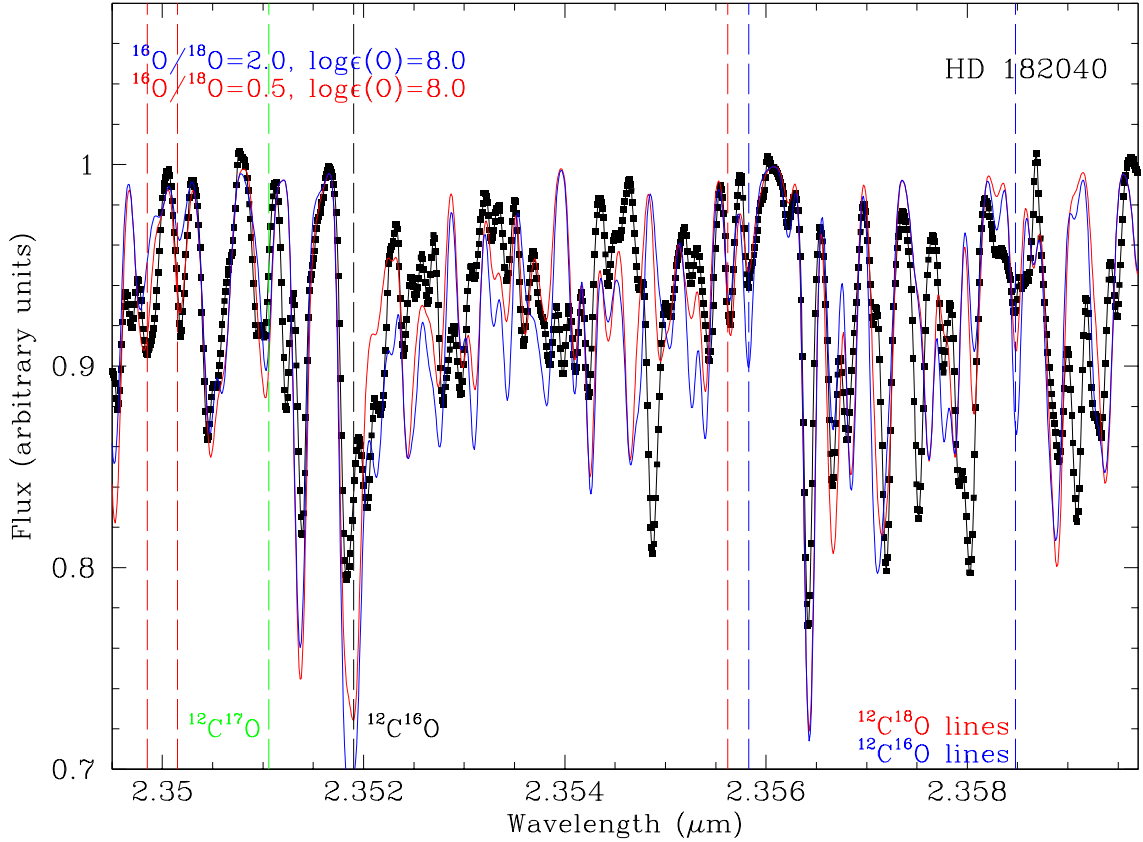}
\caption{Best synthetic (red) and observed (black) spectrum in the region around
the 2-0 2.349$\mu$m $^{12}$C$^{18}$O bandhead for the $^{18}$O-rich HdC star HD
182040. The red spectrum assumes the input C abundance of 9.52,
 the N abundance of 9.2, a
total O ($^{16}$O $+$ $^{18}$O) of log$\varepsilon$(O)=8.0 and the isotopic ratio
$^{16}$O/$^{18}$O=0.5. A synthetic spectrum for a different $^{16}$O/$^{18}$O ratio
of 2 (blue) but the same total O ($^{16}$O $+$ $^{18}$O) of log$\varepsilon$(O)=8.0
is shown for comparison. Note that there are several features (e.g. those at
$\sim$2.3540, 2.3550, 2.3575, 2.3580 and 2.3590 $\mu$m) which are not reproduced by
our synthetic spectra. \label{fig17}}
\end{figure}

\clearpage

\begin{figure}
\includegraphics[angle=-90,scale=.60]{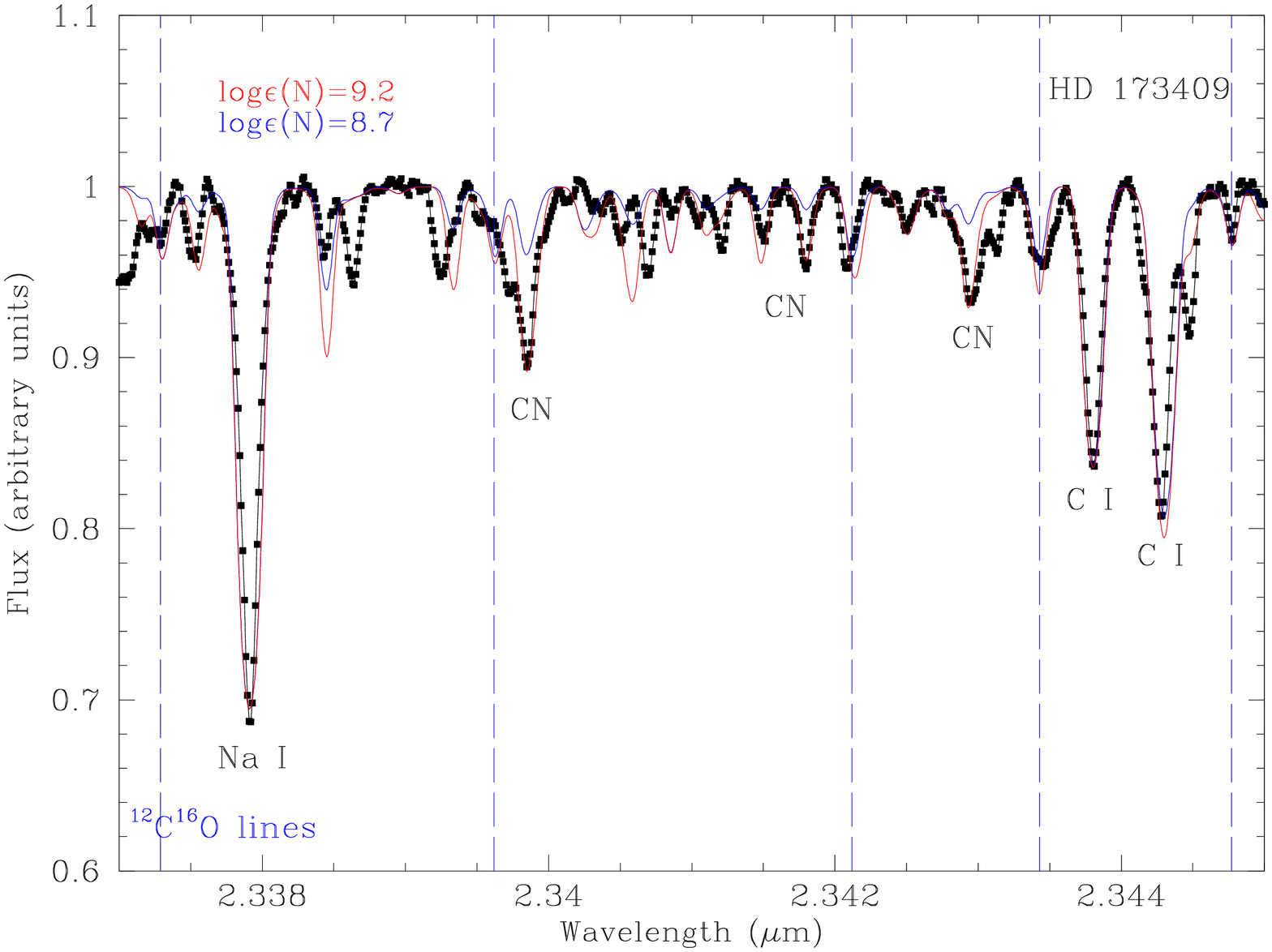}
\caption{Best synthetic (red) and observed (black) spectrum for the CO and CN
region centered at $\sim$2.341 $\mu$m for the HdC star HD 173409. The synthetic
spectrum has been created assuming the input C abundance of 9.52, the N is in
the form of $^{14}$N with an abundance log$\varepsilon$(N)=9.2 dex, and a total O
abundance fixed at log$\varepsilon$(O)= 8.7. A synthetic spectrum for a lower
$^{14}$N content of log$\varepsilon$(N)=8.7 (in blue) is also shown for
comparison.\label{fig18}}
\end{figure}

\clearpage

\begin{figure}
\includegraphics[angle=-90,scale=.60]{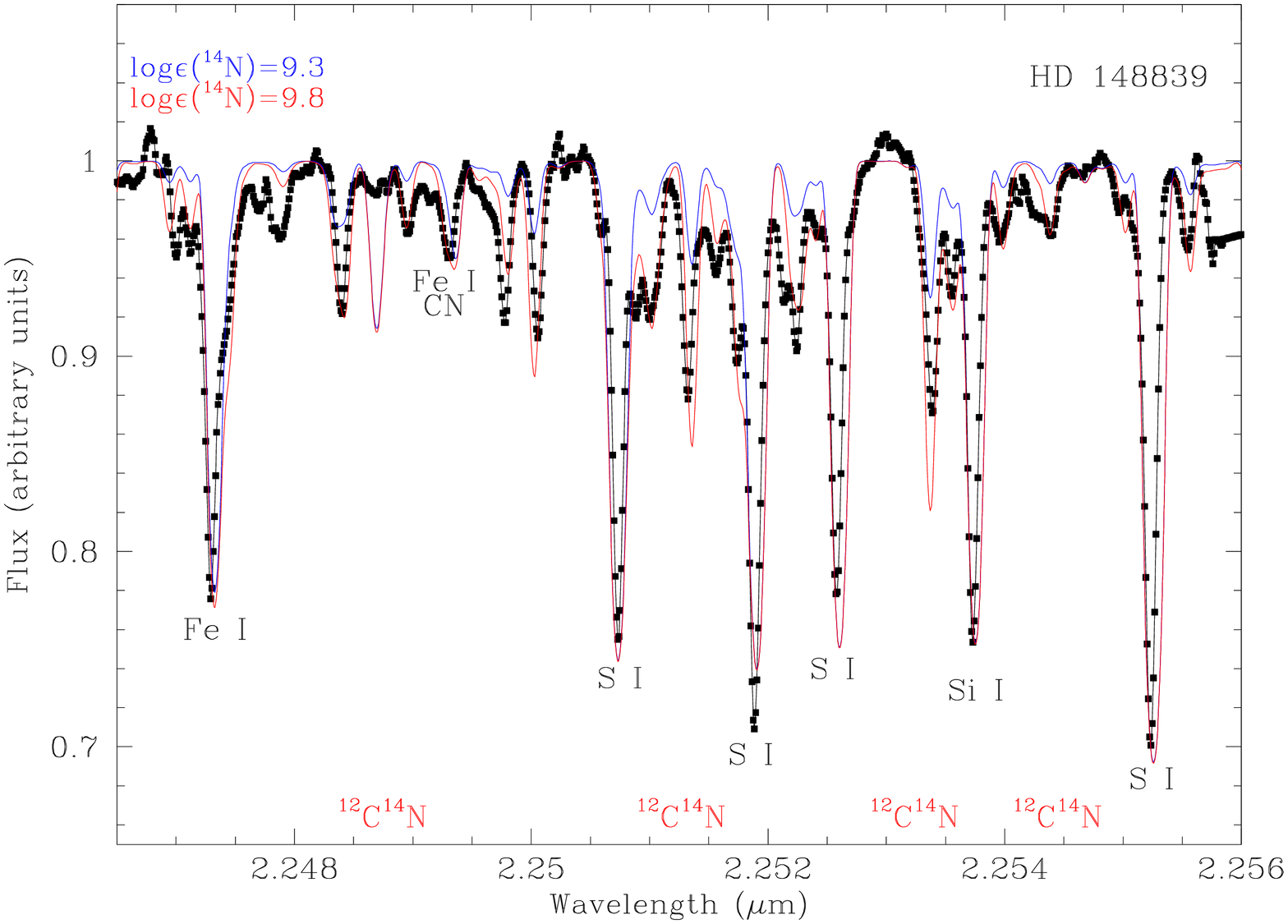}
\caption{Best synthetic (red) and observed (black) spectrum for the CN region
centered at $\sim$2.251 $\mu$m for the HdC star HD 148839. The
synthetic spectrum has been created assuming the input C abundance of 9.52,
the N is in the form of $^{14}$N with an abundance log$\varepsilon$(N)=9.8 dex,
and a total O abundance of log$\varepsilon$(O)= 9.2. A synthetic spectrum
(blue) for a lower $^{14}$N abundance of log$\varepsilon$(N)=9.3 dex is also
shown for comparison. \label{fig19}}
\end{figure}

\clearpage

\begin{figure}
\includegraphics[angle=-90,scale=.60]{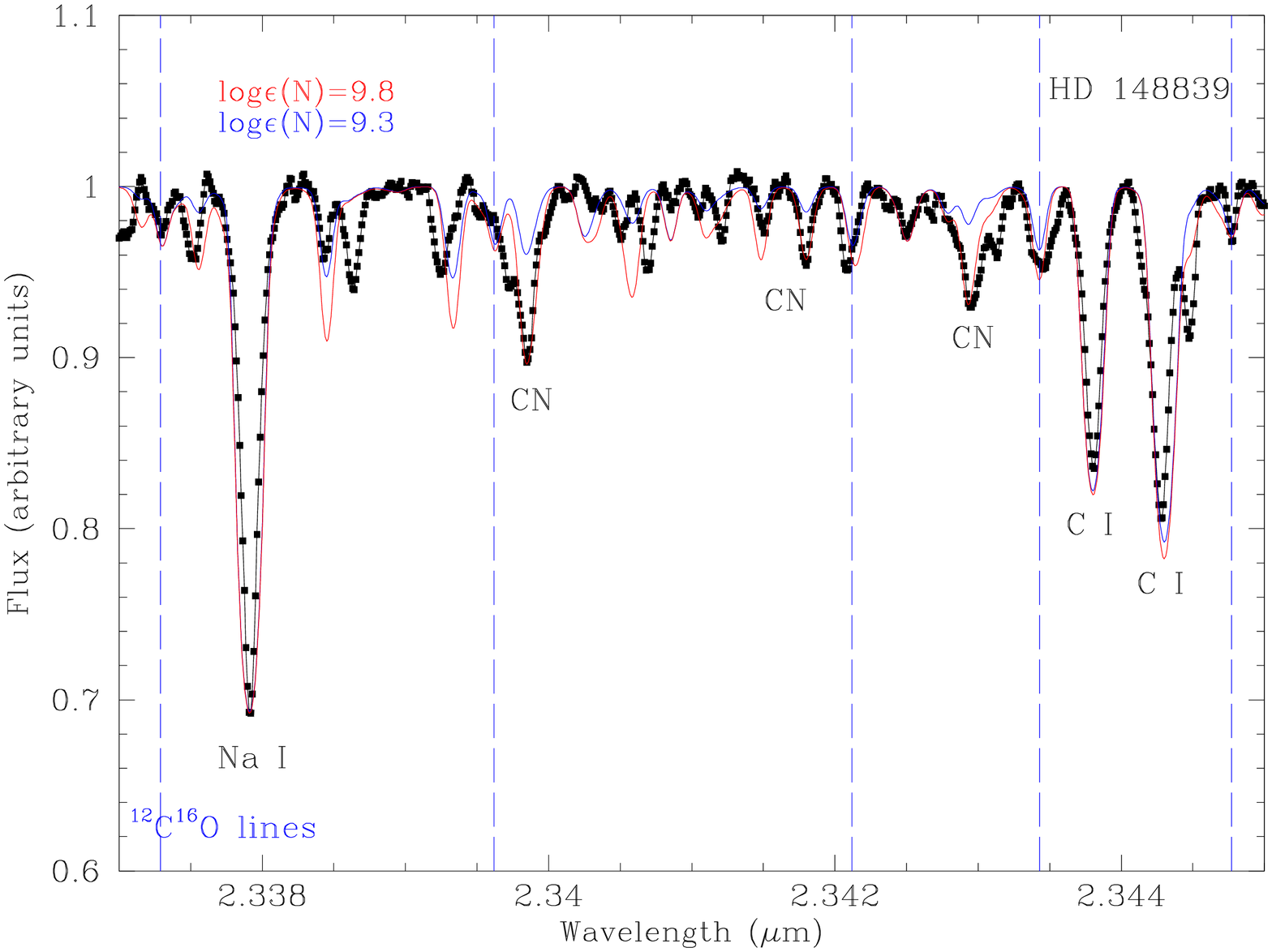}
\caption{Best synthetic (red) and observed (black) spectrum for the CO and CN
region centered at $\sim$2.341 $\mu$m for the HdC star HD 148839. The synthetic
spectrum has been created assuming the input C abundance of 9.52, the N is in
the form of $^{14}$N with an abundance log$\varepsilon$(N)=9.8 dex, and a total O
abundance of log$\varepsilon$(O)= 9.2. A synthetic spectrum for a lower
$^{14}$N abundance of log$\varepsilon$(N)=9.3 (in blue) is also shown for
comparison.\label{fig20}}
\end{figure}

\clearpage

\begin{figure}
\includegraphics[angle=-90,scale=.60]{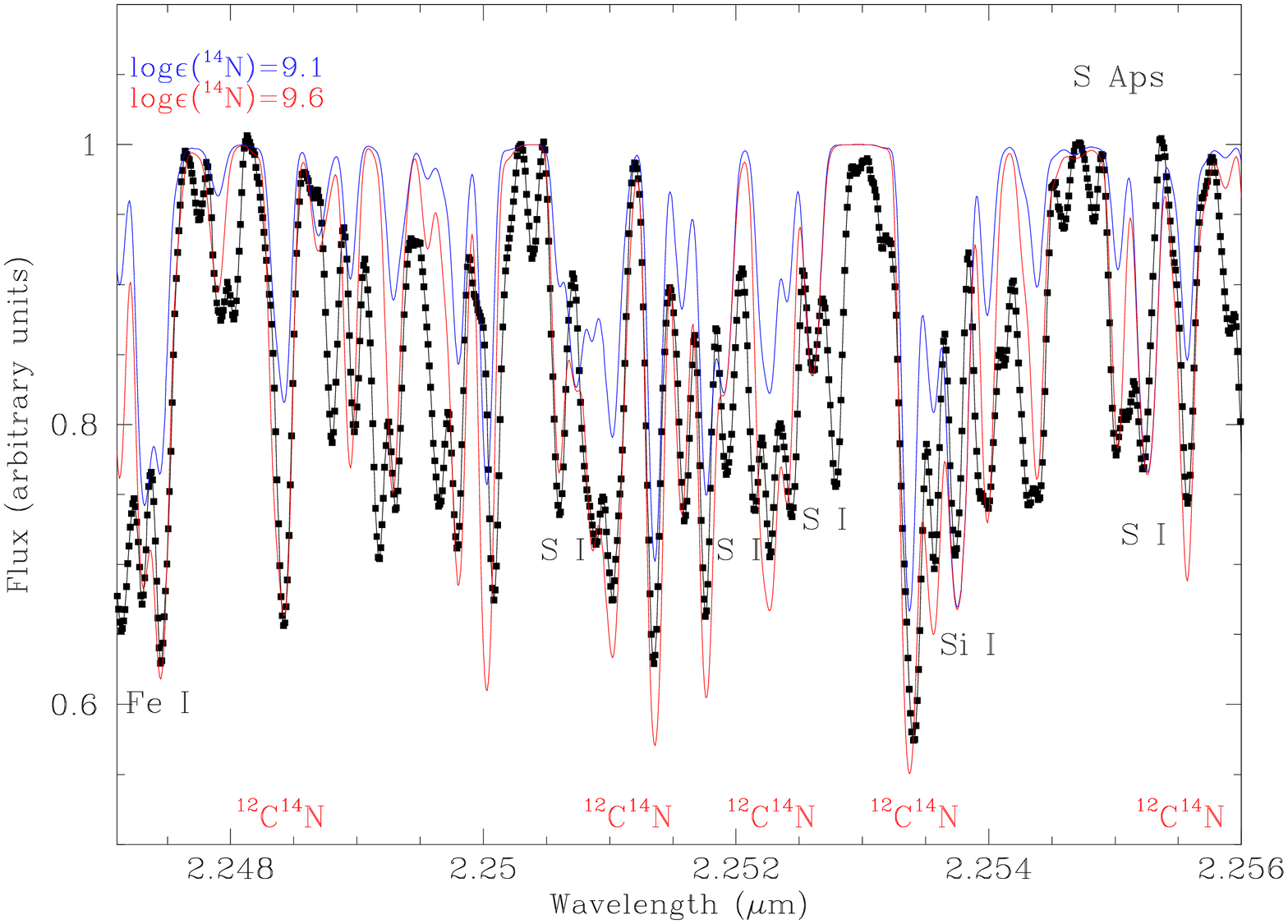}
\caption{Best synthetic (red) and observed (black) spectrum for the CN region
centered at $\sim$2.251 $\mu$m for the $^{18}$O-rich RCB star S Aps. The
synthetic spectrum has been created assuming the input C abundance of 9.52,
the N is in the form of $^{14}$N with an abundance log$\varepsilon$(N)=9.6 dex,
and a total O abundance of log$\varepsilon$(O)= 9.4. A synthetic spectrum
(blue) for a lower $^{14}$N abundance of log$\varepsilon$(N)=9.1 dex is also
shown for comparison. \label{fig21}}
\end{figure}

\clearpage

\begin{figure}
\includegraphics[angle=-90,scale=.60]{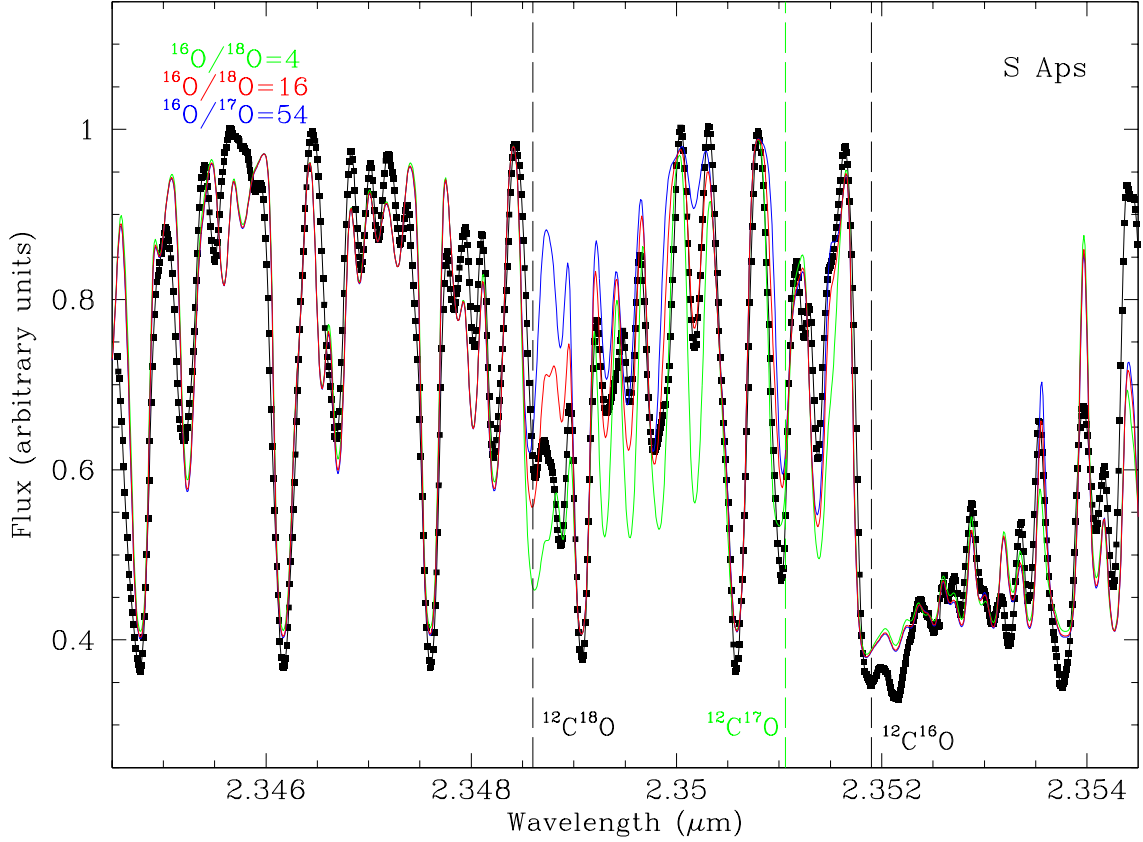}
\caption{Best synthetic (red) and observed (black) spectrum in the region around
the 2-0 2.349$\mu$m $^{12}$C$^{18}$O bandhead for the $^{18}$O-rich RCB star S Aps.
The red spectrum assumes the input C abundance of 9.52,
 the N abundance of 9.6, a total O
($^{16}$O $+$ $^{18}$O) of log$\varepsilon$(O)=9.4 and the isotopic ratio
$^{16}$O/$^{18}$O=16. Synthetic spectra for different $^{16}$O/$^{18}$O ratios of 4
(green) and 54 (blue) but the same total O ($^{16}$O $+$ $^{18}$O) of
log$\varepsilon$(O)=9.4 are shown for comparison.\label{fig22}}
\end{figure}

\clearpage

\begin{figure}
\includegraphics[angle=-90,scale=.60]{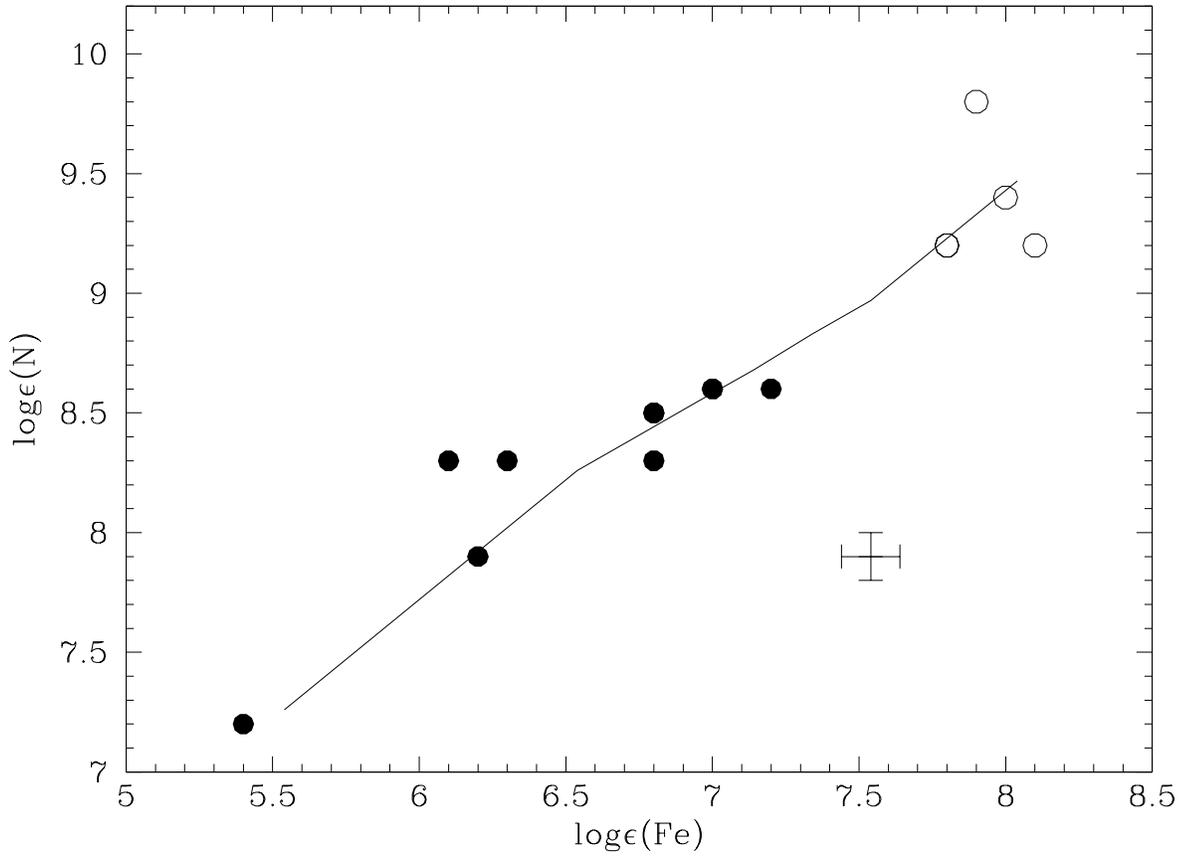}
\caption{Nitrogen abundances for EHe stars (filled circles) and HdC stars (open
circles) as a function of the Fe abundances. The solid line represents the N
abundance being equal to the sum of the initial C, N, and O abundances expected
from the Fe abundance. The solar N and Fe abundances
are indicated by a large cross. \label{fig23}}
\end{figure}

\clearpage

\begin{figure}
\includegraphics[angle=-90,scale=.60]{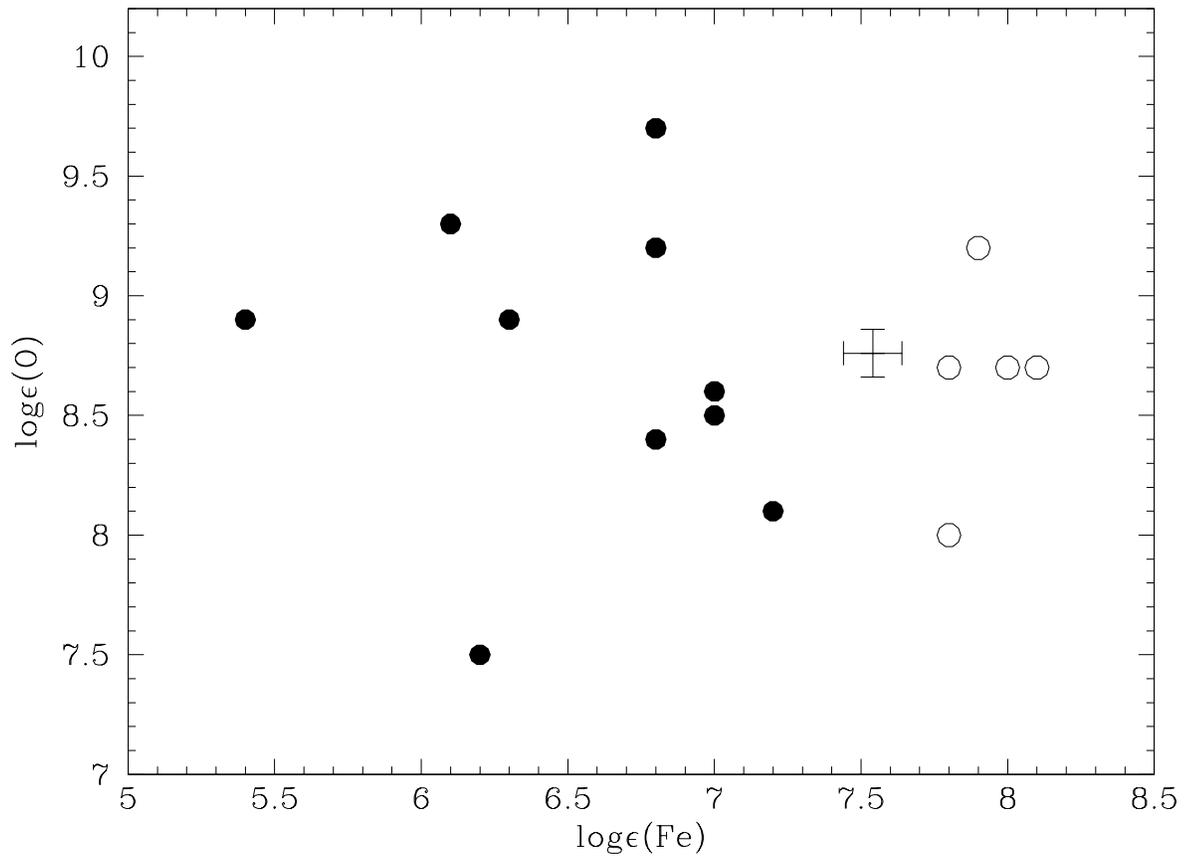}
\caption{Oxygen abundances for EHe stars (filled circles) and HdC stars (open
circles) as a function of the Fe abundances. As in Figure 23, the solar O and
Fe abundances are indicated by a large cross. \label{fig24}}
\end{figure}

\end{document}